\documentclass[letterpaper,12pt]{article}
\usepackage[affil-it]{authblk} 
\usepackage{graphicx}
\usepackage{amssymb,amsmath,amsfonts,mathrsfs,latexsym}
\usepackage{soul}
\usepackage[sort&compress, numbers]{natbib}
\usepackage{enumerate}
\usepackage{threeparttable}

\usepackage[dvipsnames]{xcolor}

\usepackage[left=1in,right=1in, top=1in, bottom=1in, headsep=.5em]{geometry}

\usepackage{amsthm}
\newtheorem{theorem}{Theorem}

\theoremstyle{definition}
\newtheorem{definition}[theorem]{Definition}
\theoremstyle{remark}

\DeclareMathOperator*{\argmin}{arg\,min}

\begin{document}

\title{Data-driven approach for synchrotron X-ray Laue microdiffraction scan analysis}
\author[1]{Yintao Song}
\affil[1]{\small{Independent Researcher, Foster City, CA, USA}}
\author[2]{Nobumichi Tamura}
\affil[2]{\small{Advanced Light Source, Lawrence Berkeley National Lab, Berkeley, CA, USA}}
\author[3]{Chenbo Zhang}
\author[3]{Mostafa Karami}
\author[3]{Xian Chen \thanks{xianchen@ust.hk}}
\affil[3]{\small{Mechanical and Aerospace Engineering, Hong Kong University of Science and Technology, Hong Kong}}
\date{}
\maketitle

\begin{abstract}
We propose a novel data-driven approach for analyzing synchrotron Laue X-ray
microdiffraction scans based on machine learning algorithms. The basic architecture
and major components of the method are formulated mathematically. We demonstrate
it through typical examples including polycrystalline BaTiO$_3$, multiphase transforming
alloys and finely twinned martensite. The computational pipeline is implemented for
beamline 12.3.2 at the Advanced Light Source, Lawrence Berkeley National Lab. The
conventional analytical pathway for X-ray diffraction scans is based on a slow pattern by
pattern crystal indexing process. This work provides a new way for analyzing X-ray
diffraction 2D patterns, independent of the indexing process, and motivates further
studies of X-ray diffraction patterns from the machine learning prospective for the
development of suitable feature extraction, clustering and labeling algorithms.
\end{abstract}

\section{Introduction}

X-ray crystallography is a fundamental tool in modern technologies for identifying or solving the
crystalline structures of solids ever since the discovery of crystal diffraction in early 1900s.\cite{maxvonlaue1912}  Today's
third generation synchrotron radiation facilities produce highly collimated and high-brilliance Xray
beams, that can be focused down to sub-micrometer sizes, opening the way to spatially resolved
quantitative studies of materials microstructures.\cite{tamura2003scanning, ulrich2011new}  Scanning Laue x-ray microdiffraction using
a pink or white x-ray beam is a technique that emerged in the late 1990s at synchrotron facilities
which has been used to map the distribution of materials structural properties, such as crystal
phase identity, crystal grain orientation, lattice distortion and degree of crystallinity, \textit{etc}. So far,
this technique has been successfully implemented at the Advanced Light Source (ALS) of the
Lawrence Berkeley National lab, the Advanced Photon Source (APS) of Argonne National Lab,
the European Synchrotron Radiation Facility (ESRF), the Canadian Light Source (CLS) and the
Taiwan Photon Source (TPS) and is in development elsewhere. This technique enables the use of
micro X-ray beam as a scanning probe to quantitatively analyze both structural and topographic
information of solid crystalline materials. \cite{tamura2003scanning}

Within the scanned area, every point illuminated by the micro-size x-ray white beam gives rise
to a diffraction pattern captured by a 2-dimensional (2D) detector with fast acquisition and short
read-out time (typically a second or less per point). The data collected by the 2D detector is a
single channel (gray-scale) image, called a \emph{Laue pattern}. The conventional way of analyzing a Laue
microdiffraction scan is to treat each pattern independently: indexing all or the majority of the
reflections in each Laue pattern with the knowledge of the crystal structure (space group, atomic
types and unique positions within unit cell and lattice parameters). By performing the indexation
for all the Laue patterns in a scan, the crystal orientation (or grain structure) distribution of the
material can be obtained and displayed as quantified color maps.

The success of the crystallographic analysis for a Laue microdiffraction scan strongly depends
on the indexing result of individual Laue patterns. For cubic structure, the indexing procedure is
straightforward if the space group and Wyckoff positions\cite{wyckoff1922analytical} are known. Due to the nature of white
beam diffraction, the lattice parameter absolute values do not contribute much to the indexing of
reflections diffracted by the cubic structure (for instance, two fcc crystals with comparable lattice
parameters and identical orientation will give nearly identical Laue patterns). However, for crystal
structures with symmetry lower than cubic, the relative sizes and angles of the unit cell 
({\it i.e.} $a/b$, $a/c$, $\alpha$, $\beta$ and $\gamma$) play an important role in the indexing procedure. For these crystals, many
uncertainties can arise in the indexing results when the lattice parameters are not well known and
at least one of the lattice parameters is fairly large. Slight perturbations of the lattice parameters might result in different indices
corresponding to the same reflection in the Laue pattern, {\it i.e.} misindexation. Sometimes, the
indexing procedure would fail when the initial guess of the lattice parameters is far from the true
values for the tested material.

The outcome of modern X-ray microdiffraction experiments are data-rich. A typical 2D scan
generates from thousands to hundred thousands diffraction patterns. From the computational
point of view, the iterative indexing calculations of individual Laue patterns are often redundant
and time expensive for large datasets. The analysis and indexing results of all Laue patterns
within an iso-oriented spatial domain (a crystal grain) are almost identical. Using crystallographic
analysis tools such as XMAS (X-ray microdiffraction analysis software) \cite{tamura2014xmas}, the iso-oriented regions
with specific color labels are identified. However, when these tools are applied to label a large area
comprised of thousands of hundreds of Laue patterns, it usually takes days to finish the calculation
by the optimized indexing algorithm running on a desktop computer, and synchrotron facilities
have now opted to use parallel versions of the indexing code running on powerful computational
GPU or CPU clusters. However, scientists who collect data at the beamline do not necessarily
have access to these clusters at their home institution. If the scanned domain comprises of multiple
phases with different orientations, the complexity of the analysis and the computational time will
dramatically increase. For instance, the processing of the BaTiO$_3$ scan mentioned below (Figure
\ref{fig:BTO}) and consisting of 6000 Laue patterns take about 15 minutes using 600 nodes on a cluster, but
the analysis would have taken 6 days on a regular desktop machine.

From another perspective, the analysis of Laue scans is image processing. One of the key steps in the whole procedure is to recognize the similarities between the gray-scale Laue patterns. Therefore, it is nature to introduce the deep learning methods to assist the analysis. Along the way of algorithmic development for imaging processing, the convolutional neutral network have been widely introduced for medical image analysis \cite{litjens2017survey} and computed tomography reconstruction \cite{schluter2014ct}. Recent years, the machine learning algorithms have been developed for structural determination of X-ray powder diffraction \cite{park2017iucrj} and X-ray protein crystallization image \cite{yann2016crystalnet}. Similar concepts have also been applied to the electron micrographs for defect inspection \cite{liu2008image},  crystal recognition \cite{li2018automated} as well as the diffraction based experiments \cite{xu2018deep, jha2018extracting}. In these applications, the images for training and the object for segmentation / analyzing are usually of the same type. For example, they are both CT images or both X-ray diffraction spectra. In contrast, in Laue X-ray microdiffraction experiments, the latent features are extracted from the diffraction patterns (i.e. a set of gray-scale $\sim$ 1M pixel images), while the experimental domain is a 2D meshgrid consisting of a series of Laue patterns occupying certain spatial domain on sample surface. For this reason, the existing deep learning feature extractors for image segmentation \cite{fauvel2007, plaza2009} is not very suitable for microdiffraction analysis. The goal of the experiment is to deliver the orientation ``contrast" among various sub-domains within the scanned area. 

In this paper, we propose a data-driven approach to abate the burden of indexing calculation
for the analysis of synchrotron Laue X-ray microdiffraction scans so that analysis can be performed
quickly without the need of a supercomputer. The main goal is to outline the methodology of the computational pipeline with the application to the data segmentation for Laue X-ray
microdiffraction experiment by machine learning algorithms. Our pipeline is comprised of two main steps: 1) feature extraction of Laue patterns; 2) classification and labelling for all sequential Laue patterns collected in the spatial domain. Our pipeline has been implemented
on beamline 12.3.2 at the Advanced Light Source, Lawrence Berkeley National Lab, but can be
easily extended to similar X-ray diffraction experiments at any facility.



\section{Methodology}\label{sec:description}

\begin{figure}
\centering
\caption{Orientation map of a BaTiO$_3$ specimen from complete indexing. Scanning step size is $5\mu{\rm m}$ in both directions.}\label{fig:BTO}
\includegraphics[width=3in]{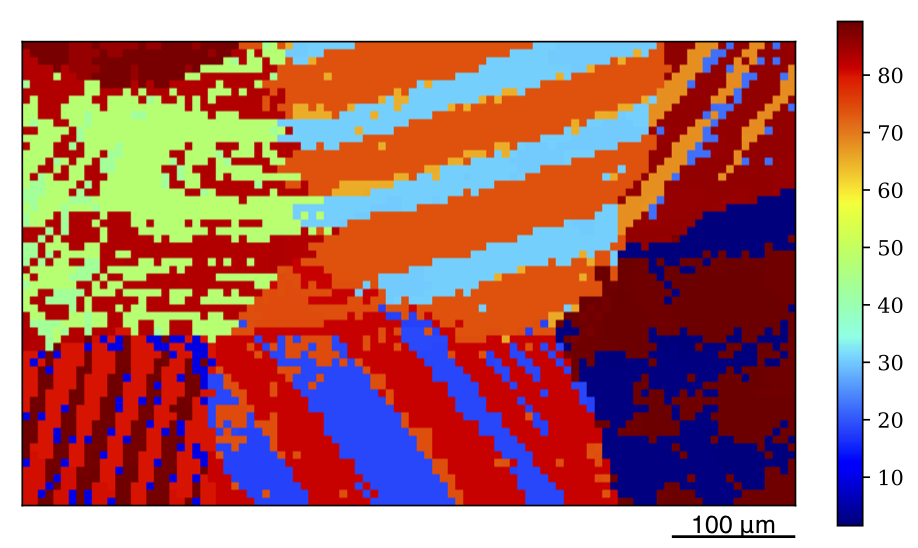}
\end{figure}

Our idea originates from the nature of X-ray microdiffraction results. The spatial distribution of
a certain property is usually a piecewise constant function. For example, Figure \ref{fig:BTO} shows the orientation
map of a martensite polycrystal BaTiO$_3$ represented by the angle between the a crystal
axis and the normal to the sample surface. The regions of same color correspond to the same
crystallographic orientation up to small variations. The color map is calculated and generated by
XMAS \cite{tamura2014xmas}, revealing the orientation distribution, morphology of grain boundaries and microstructure
within each of the grains. Given the unstrained lattice parameters and the stiffness tensor
of the tested material, the orientation map can be converted into a strain or stress map.\cite{chen2016quantitative}  
The piecewise scalar/vectorized color values of these maps represent certain physical property of the
material. Each of the color values corresponds to a specific crystallographic indexing. Within each
iso-oriented region (corresponding to an iso-indexing cluster), only one delegate Laue pattern needs
to be chosen for crystallographic analysis. The indexing of the rest of the patterns in the cluster is
then automatically determined.

\begin{figure}
\centering
\includegraphics[width=2.5in]{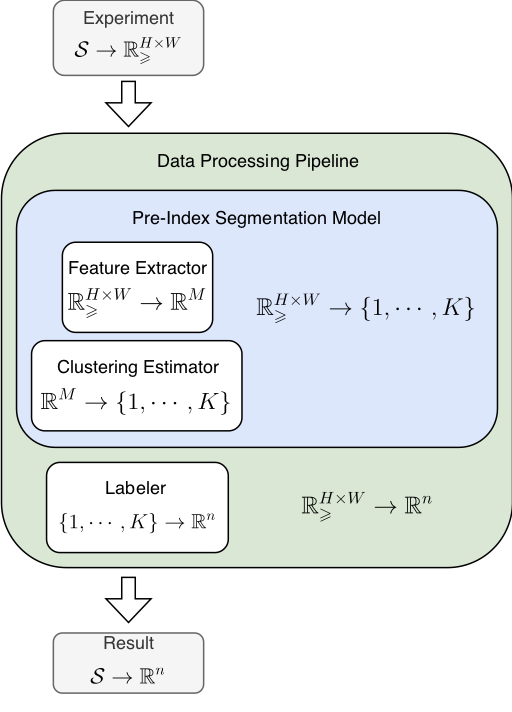}
\caption{Data-driven micro Laue XRD data processing pipeline.}
\label{fig:pipeline}
\end{figure}

In this section, we propose an indexing-free data segmentation method for X-ray microdiffraction
scans. The overall process pipeline is illustrated in Figure \ref{fig:pipeline}. In principle, an iso-indexing
cluster should consist of similar Laue patterns. From the experiment, we obtain a set of singlechannel
images, from which a learning model can be designed to extract the features to identify
their similarities. By using a machine learning algorithm, these images are divided into a set of
clusters based on the extracted features. Then the inverse map of the clusters in feature space
naturally forms the pre-index segmentation in spatial domain of the scan.

In order to clearly describe the clustering and labeling methods and algorithms in our approach,
we provide the following formal definitions for the X-ray Laue microdiffraction experiment.

\begin{definition}
The scanned area on the specimen is called the \emph{specimen domain} $\cal S$.
It is represented by a 2D mesh grid $\{1,\cdots,N_x\}\times\{1,\cdots,N_y\} \subset {\mathbb N}^2$.
$N_x$ and $N_y$ are the number of steps along $x$ and $y$ direction respectively. 
The total number of grid points (scanned locations) is $N = N_x N_y$.
\end{definition}
The step size along either directions is a small real number underlying the resolution of the Laue microdiffraction. Regarding the data segmentation, the values of step sizes along $x$ and $y$ directions do not affect the clustering result. Therefore, the specimen domain is considered as a 2D integer domain.
\begin{definition}
A \emph{Laue pattern}, or simply a \emph{pattern}, at a grid point $(x, y)$ is a $H \times W$ gray-scale image $I_{x, y} \in \mathbb R_{\geqslant}^{H \times W}$. 
A \emph{Laue diffraction experiment}, or simply a \emph{experiment} is a mapping ${\cal I}:{\cal S} \to {\mathbb R_{\geqslant}^{H \times W}}.$
\end{definition}

\begin{definition}\label{def:property-map}
A \emph{property map} on the specimen domain $\cal S$, is a function 
$f:{\cal S} \to \mathbb R^n$. $n$ is the dimension of the property.
\end{definition}
An experiment itself is a property map of the dimension $H \times W$.
But it is not an interesting one.
In practice, the goal is almost always to find a property map with some physical significance, such as the orientation of the crystal.
The procedure being introduced in this paper is going to find an approximated map that is close to the true map but requires much less, if any, indexing effort.

\begin{definition}\label{def:feature-map}
A \emph{feature extractor} $V$ maps a diffraction pattern $I$ to a $M$ dimensional feature vector $\mathbf v$.
That is 
\begin{equation}\label{eq:feature-extractor}
    V: {\mathbb R}_{\geqslant}^{H \times W} \to \mathbb R^{M}.
\end{equation}
${\mathbb R}^M$ is the \emph{feature space}.
$M$ is also called the \emph{number of features}.
Combined with an experiment ${\cal I}$, we get the \emph{feature map} $v = V\circ{\cal I}$.
The map
\begin{equation}\label{eq:featuremap}
    v:{\cal S} \to \mathbb R^M
\end{equation}
gives the feature vector $v(x, y)$ at each grid point $(x, y)$.
The feature vectors in the image $v({\cal S}) \subset \mathbb R^M$ are called the \emph{feature samples}.
\end{definition}
Usually, $M \ll H\times W$ in practice, which is the motivation of extracting features out of diffraction patterns.
\begin{definition}\label{def:feature-transformation}
A \emph{feature transformation} $T$ maps a feature vector $\mathbf v_1\in {\mathbb R}^{M_1}$ to another feature vector $\mathbf v_2\in {\mathbb R}^{M_2}$.
{\it i.e.} $T:\mathbb R^{M_1} \to \mathbb R^{M_2}$.
\end{definition}
A straightforward collary from the above definition is that the composition $\tilde V$ and $\tilde v$
\begin{equation}
\begin{cases}
    & \tilde V = T_n \circ \cdots \circ T_i \circ \cdots \circ T_1 \circ V \\
    & \tilde v = T_n \circ \cdots \circ T_i \circ \cdots \circ T_1 \circ v \\
    & V:{\mathbb R}_{\geqslant}^{H \times W} \to \mathbb R^{M_0} \\
    & v:{\cal S} \to \mathbb R^{M_0} \\
    & T_i:\mathbb R^{M_{i-1}} \to \mathbb R^{M_i}
\end{cases}
\end{equation}
are still a feature extractor and a feature map to $M_n$ features.

\begin{definition}
A \emph{clustering estimator}, or simply an \emph{estimator}, $g$, of $K$ clusters in the feature space $\mathbb R^M$ is a map
\begin{equation}
   g:{\mathbb R}^M \to \{1,\cdots,K\}.
\end{equation}
The integer $K$ is the \emph{number of clusters}.
For each $k \in \{1,\cdots,K\}$, the \emph{cluster} ${\cal C}_k \subset \mathbb R^M$ is the equivalent class
\begin{equation}
    {\cal C}_k = \left\lbrace \mathbf v \in {\mathbb R}^M : g(\mathbf v) = k  \right\rbrace.
\end{equation}
If the feature space is extracted by the feature map $v$, the specimen domain $\cal S$ is partitioned into \emph{subdomains}
\begin{equation}
    {\cal S}[v]_k = \left\lbrace (x, y) \in {\cal S} : v(x, y) \in {\cal C}_k  \right\rbrace.
\end{equation}
When there is no confusion about $v$, we ignore the parameter $v$, and simply write  ${\cal S}_k$.
\end{definition}

The motivation of partitioning feature spaces into clusters is the belief that the property of interest is (almost) constant across all feature vectors in one cluster.
We call this shared property the \emph{label} of a cluster. 
\begin{definition}
A \emph{labeler} $\ell$ with $n$ \emph{channels} of $K$ clusters is the map
\begin{equation}
    \ell : \{1,\cdots,K\} \to {\mathbb R}^n
\end{equation}
$\ell(k)$ is the \emph{label} of cluster ${\cal C}_k$.
\end{definition}
Again, recall Definition \ref{def:property-map}, $n$ is the dimension of the shared property.

Clearly, the identity mapping $\ell_{\rm N}(k) = k$ is a labeler for any number of clusters.
We call $\ell_{\rm N}$ the \emph{natural labeler}, the resulting label $k$ is the \emph{natural label} of ${\cal C}_k$.

\begin{definition}\label{def:pipeline}
A triplet $(v, g, \ell)$ of (feature map, estimator, labeler) is called a \emph{data processing pipeline} or simply \emph{pipeline}. 
The first two phases $(v, g)$ is called a \emph{pre-index segmentation model}, or simply \emph{segmentation model} or \emph{segmentation}.
\end{definition}

\begin{definition}\label{def:clustered-property}
For a pipeline $(v, g, \ell)$, the property map defined as
\begin{equation}
    \phi[v, g, \ell] = \ell \circ g \circ v.
\end{equation}
is called the \emph{label map} generated by the pipeline.
\end{definition}
The label map in Definition \ref{def:clustered-property}, 
$\phi$ is map from spatial domain $\mathcal S$ to the space $\mathbb R^n$.
This map is the approximation of the true property map from the complete indexing. 
By definition, this approximated map is determined by the pipeline $(v, g, \ell)$.
In following sections, we are going to discuss the strategies of constructing each phase of the pipeline.

\begin{definition}\label{def:centroid}
A \emph{centroid assignment} $u$ assigns a feature vector to each cluster:
\begin{equation}
    u:  \{1,\cdots,K\} \to {\mathbf u}_k.
\end{equation}
The mapped vector $\mathbf u_k$ is called \emph{centroid} of the cluster ${\cal C}_k$.
\end{definition}

The exact centroid assignment depends on the clustering algorithm.
Normally it is the mean position of all the points (in the training set) in a cluster.
Note that the centroid of a cluster may not be a feature sample (Definition \ref{def:feature-map}).
For example, the centroid can be the mean of all feature vectors in a cluster.
Because of this, we need to introduce the concept of \emph{delegate} samples.
\begin{definition}\label{def:delegate }
The \emph{delegate  point} $(x_k, y_k)$ of the cluster ${\cal C}_k$ is given by a \emph{delegate  assignment}
\begin{equation}
    w: \{1,\cdots,K\} \to {\cal S}.
\end{equation}
The \emph{delegate  sample} (or the \emph{delegate  feature vector}) of ${\cal C}_k$ is
\begin{equation}
    \mathbf w_k = v(x_k, y_k).
\end{equation}
\end{definition}
Usually we pick the feature sample closest or equal to the centroid:
\begin{equation} \label{eq:delegate}
    (x_k, y_k) = \argmin_{(x, y)\in{\cal S}_k \cap \{(x, y): I_{x, y}\text{ can be indexed}\}} \left\Vert v(x, y) - {\mathbf u}_k \right\Vert
\end{equation}
The assignment $w(k)$ from the selected delegate sample calculated by equation \eqref{eq:delegate} might not exists if none of the patterns in the cluster can be indexed. 
Note that both the centroid assignment $u$ and the delegate  assignment $w$ are labelers of $K$ clusters.

The main purposes of labeling are two-fold: 1) Reduce the feature space of dimension $M$ to
a lower dimensional space $\mathbb R^n$ such that its distribution over the specimen domain $\mathcal S$ can be well
presented and visualized. 2) Since the label itself is also a feature derived from the original Laue
patterns, the label map creates a spatial distribution of such a derived feature. In general, this
derived feature does not have any physical meaning, but it quantifies the similarities between Laue
patterns. Thus, we need to index the delegate pattern for each cluster, to associate each of the
label values, and therefore to create the whole label map with certain physical meaning. However,
if we use a labeler that is closely related to some physical properties, then the label map is itself
a distribution with physical significance. Based on the aforementioned methodology, the feature
extraction and clustering of the Laue patterns are independent of the indexing process. Figure \ref{fig:pipeline}
sketches the general procedures of a data processing pipeline.

\section{Feature extraction and transformation}

In this section and the next, we are going to walk through the procedure of constructing the
data processing pipeline, using a dataset from a polycrystalline multi-phase BaTiO$_3$ sample as
an example. The orientation map as a result of the complete indexing shown in Figure \ref{fig:BTO} is the
reference map, by which the label map generated by our data-driven approach will be assessed.

\subsection{CNN Autoencoder}

Original Laue patterns in our examples are images of about 1M pixels. If we use all pixels as
features, the number of dimensions become too high. Thus we use a Convolutional Neural Network
(CNN) autoencoder \cite{hinton2006reducing} to reduce it into a manageable number of dimensions, called \emph{latent features}.

An autoencoder is a dimension reduction technique for unlabeled data that consists of an
encoder and a decoder. The encoder compresses each high dimensional data into a low dimensional
vector, $i.e.$ latent features, then the decoder, normally with a mirrored architecture compared with
the encoder, inflates the low dimensional vector back to the reconstructed data in the original
dimension. The autoencoder is trained by minimizing a certain distance function between the
original and reconstructed data. \cite{rumelhart1986learning} A CNN autoencoder uses a pair of mirrored CNNs as the
encoder and decoder.

\begin{figure}
\centering
\includegraphics[width=3.3in]{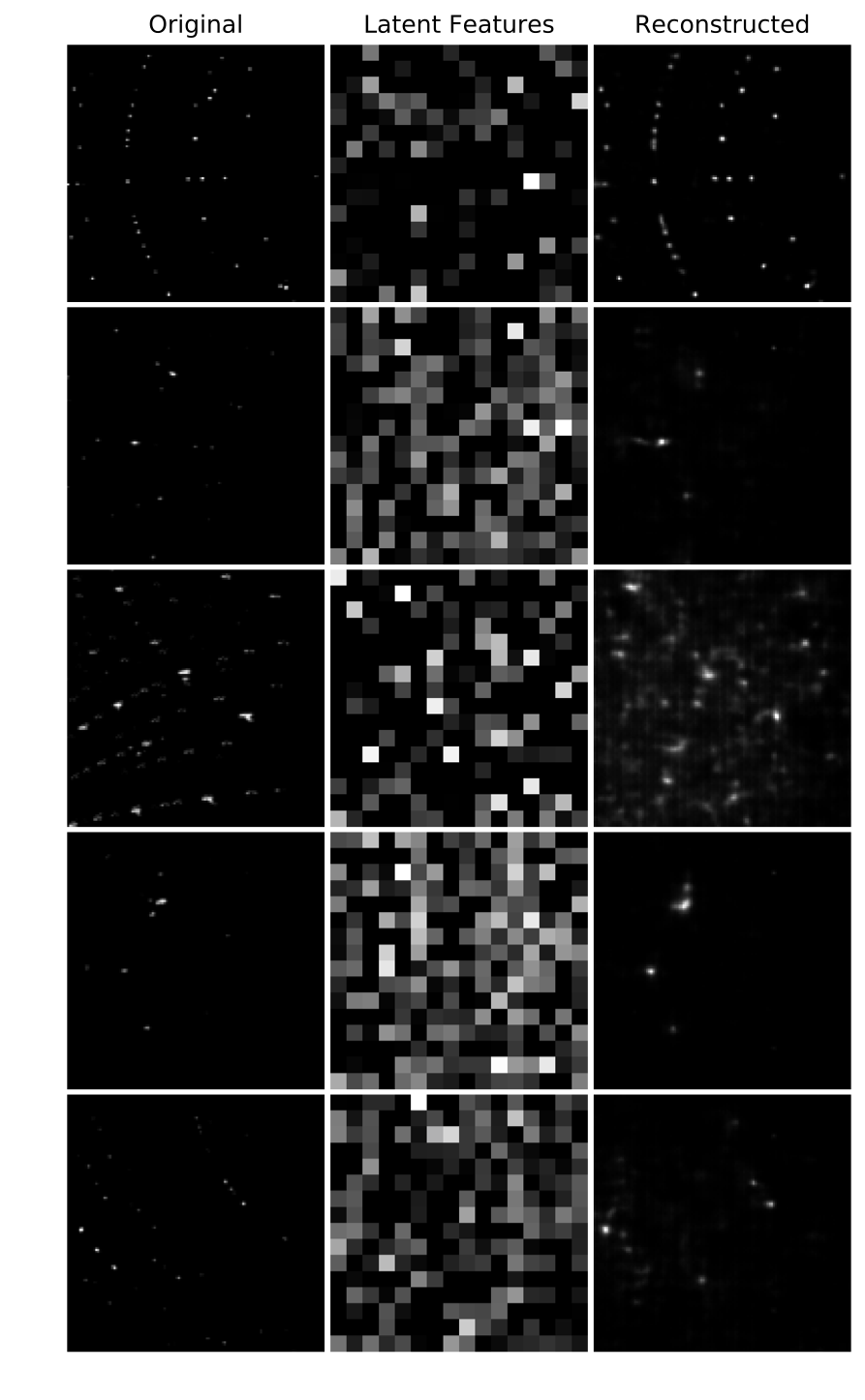}
\caption{Latent features extraced by autoencoder ($M_{\rm AE} = 256$).}
\label{fig:autoencoder}
\end{figure}

First, we crop the original images to a square shape and then shrink them to $128\times128$ pixels.
As seen in Figure \ref{fig:autoencoder}, a Laue pattern is an overall black image with a few sparsely distributed high
intensities regions called ``peaks" or ``reflections". For the purpose of segmentation, the exact peak
profile is not as important as the peak positions and the symmetric distribution of these peaks in
a Laue pattern. Therefore using a smaller image size can not only speed up training, but more
importantly the encoding for future use after training.

\begin{figure}
\centering
\includegraphics[width=0.9\textwidth]{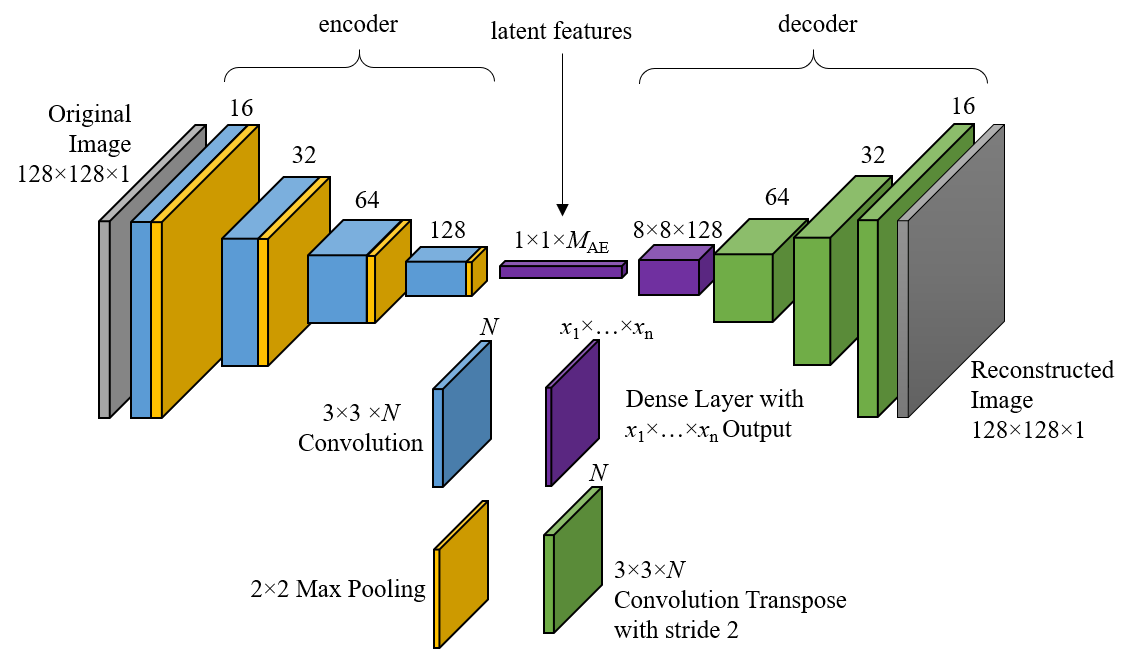}
\caption{Architecture of the CNN autoencoder.}
\label{fig:architecture}
\end{figure}

The architecture of our CNN autoencoder is illustrated in Figure \ref{fig:architecture}. 
Dropout layers \cite{srivastava14a} with dropping rate 20\% are inserted after max pooling layers and before convolution transpose layers and before the final reconstruction layer to reduce overfitting, which are omitted in the picture. 
The bottleneck layer of shape $1 \times 1 \times M_{\rm AE}$ represents the $M_{\rm AE}$ latent features.

\begin{figure}
\centering
\includegraphics[width=0.8\textwidth]{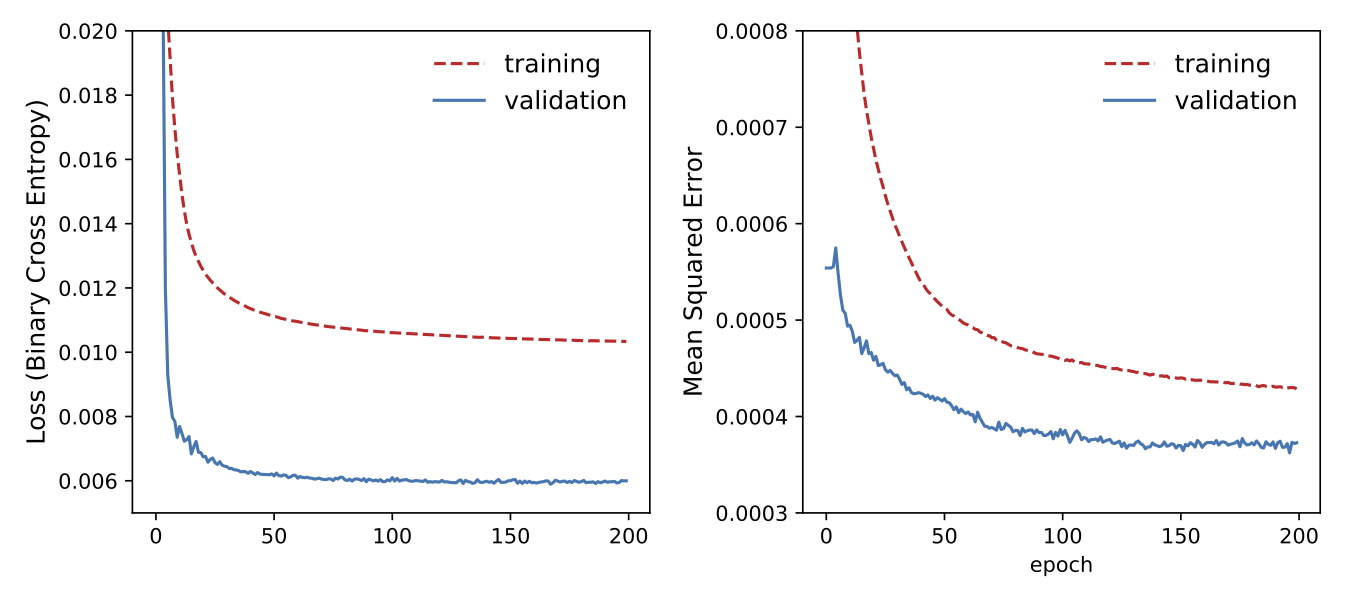}
\caption{Binary cross entory as the loss function (left) and the mean squared error 
(right) at the end of each training epoch.}
\label{fig:history}
\end{figure}

We train a generic autoencoder,
using Keras (v2.2.4) \cite{chollet2015keras} backed by TensorFlow (v1.13.1) \cite{tensorflow2015-whitepaper}, that can distinguish the differences between any two general
Laue patterns, rather than only those in one particular experiment. The training set consists of 50,000
patterns collected from over 100 different and independent experiments, and the validation set has 100
patterns carefully chosen from a different set of experiments to cover as many different types of Laue patterns as possible. 
The set of training data covers the experimental Laue patterns including polycrystalline ceramics of different symmetries, Cu, Al, Ni, Ti superalloys, shape memory alloys in austenite and martensite, Fe based alloys and inter-metallic compounds, monochromatic diffraction, quartz, deformed Nitinol martensite, Heusler alloys and so on. 
None of the experiments used in the training set is in the examples of demonstration.
Rectified Linear Units (ReLU) \cite{agarap2018deep} are used as the non-linear activation of neurons, except the final reconstruction layer, which uses the sigmoid function.
We find that directly using mean squared error as loss function traps the optimizer in the local minimum corresponding to reconstructing a completely black image.
Thus, we use the binary cross entropy as the loss function for training, and monitor the mean squared error at the end of each epoch.
The chosen optimization algorithm is Adam \cite{kingma2014adam}.
The training history is shown in Figure \ref{fig:history}.
The final model used in following examples has binary cross entropy 0.0059 and mean squared error 0.00037, evaluated on the validation set.

Encoding examples from one of the autoencoder we trained are shown
in Figure \ref{fig:autoencoder}. The difference between the original patterns can be captured reasonably well by the
difference in latent features.
We also perform some hyper-parameter tuning, including the number of hidden features and layers, the dropout rate, adding/removing dense layers, and different optimization algorithms. 
The chosen model performs slightly better than others, and, as seen in later sections, is adequate for the our demonstration. 

Using the notations introduced in Section \ref{sec:description}, we have the autoencoder as a generic feature extractor 
\begin{equation}
    V_{\rm AE}: [0, 1]^{128 \times 128} \to \mathbb R^{M_{\rm AE}}.
\end{equation}
This extractor is general, which can be used for any experiments when 
an image pre-processing is used to convert the size and intensity to $[0, 1]^{128 \times 128}$.
Therefore, for any experiment, a feature map parameterized by $M_{\rm AE}$ can be generated as
\begin{equation}\label{eq:vAE}
    v_{\rm AE}: {\cal S} \to \mathbb R^{M_{\rm AE}}.
\end{equation}

\subsection{PCA transformation}

Principal component analysis (PCA) is an orthogonal projection of feature space to another vector
space which is often of lower dimension than the original feature space, such that the variance
of the projected feature samples is maximized. \cite{bishop2006pattern, jolliffe2002pca} Each axis of the projected orthogonal basis
is called a \emph{principal component}. The variance of the projected feature samples along the $i$-th axis
is called the explained variance of the $i$-th principal component. Without loss of generality, it is a
common practice to sort the projected orthogonal basis, or the principal components, in descending
order of their explained variance.

We always apply a PCA transformation to the result of CNN autoencoder.
So the input dimension is $M_{\rm AE}$, the output dimension is $M_{\rm PCA} \leqslant M_{\rm AE}$.
$M_{\rm PCA} < M_{\rm AE}$ means the feature space is truncated to the first $M_{\rm PCA}$ principle components. 
Let 
\begin{equation}
    T_{\rm PCA}:{\mathbb R}^{M_{\rm AE}} \to {\mathbb R}^{M_{\rm PCA}}
\end{equation}
be the PCA transformation, then the composition of it and the CNN autoencoder defines
the feature map 
\begin{equation}\label{eq:v_pca}
    v_{\rm PCA} = T_{\rm PCA} \circ v_{\rm AE}: {\cal S} \to \mathbb R^{M_{\rm PCA}}.
\end{equation}

\begin{figure}
\centering
\includegraphics[width=0.8\textwidth]{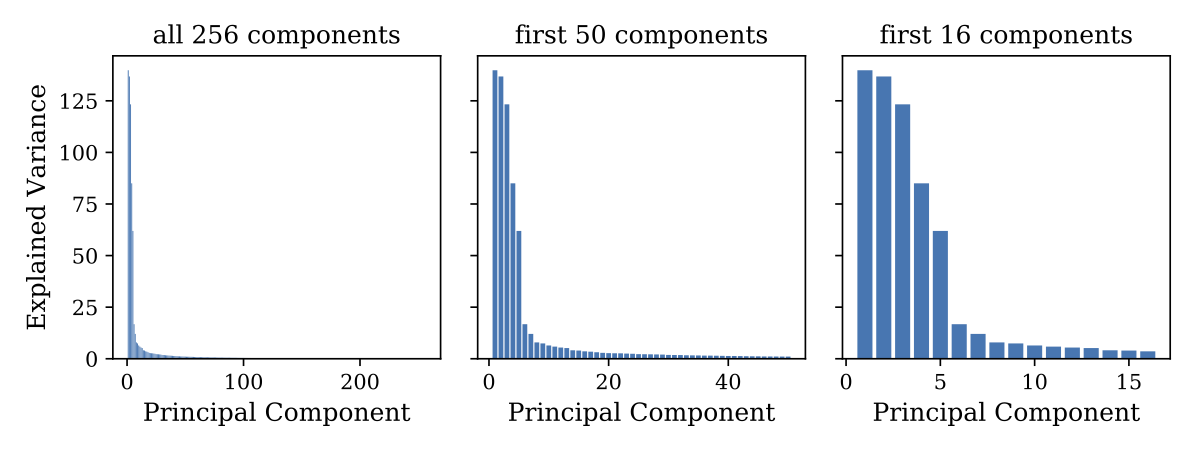}
\caption{Explained Variance of Principal Components in BaTiO$_3$.}
\label{fig:pca}
\end{figure}

For our study case BaTiO$_3$, we apply PCA to the latent features extracted by an autoencoder
with $M_{\rm AE}=256$, then plot the explained variance for each principal component as shown in Figure
\ref{fig:pca}. Clearly, the distribution is so highly skewed that the majority of weight only concentrates in
the first few components. This suggests that the truncation of the feature dimensions to the first
few principal components does not affect much of the clustering accuracy for the composed feature
map given in \eqref{eq:v_pca}. This PCA truncation will be discussed in more details in the next section.

\section{Clustering and labeling}

After mapping each of the patterns into a feature vector, we can use clustering algorithms to train
an estimator g. We consider two clustering techniques, namely K-Means \cite{bishop2006pattern, lloyd1982least} and Bayesian
Gaussian Mixture (BGM) (also known as Variational Mixture of Gaussians) \cite{bishop2006pattern, attias1999inferring}. K-Means is a
fast algorithm suitable for clusters that are well separated. It is scalable to huge data sets. The
number of clusters $K$ is crucial and mandatory for a K-Means estimator. BGM on the other hand is
a more advanced clustering that fits ellipsoidal clusters better than K-Means (which only assumes
spherical clusters). With the usage of variational Bayesian inference, BGM is also more robust on
the number of clusters. \cite{blei2006variational} BGM assigns a weight to every cluster. 
The number of clusters with nonnegligible
weight (effective clusters) is smaller than the nominal number of clusters, and is stable
as the nominal number of clusters increases. As a trade off, BGM is much more computationally
expensive compared to K-Means. It scales poorly with the size of the data set.

According to our notation, a K-Means estimator and a BGM estimator with parameter $K$ are denoted as
$g_{\rm KM}$ and $g_{\rm BGM}$ respectively.
Using either one of the two clustering algorithms combined with the feature map in \eqref{eq:v_pca}, we obtain a pre-index segmentation model (Definition \ref{def:pipeline}).
Such a segmentation has parameters $\{M_{\rm AE}, M_{\rm PCA}, K, \{KM, BGM\}\}$.
For common machine learning applications, the selection of each of the parameters known as the \emph{model selection} relies on the \emph{metric analysis}: define and understand a metric that tells the ``goodness'' of a segmentation.
In the following part, we are going to study model selection for supervised and unsupervised labeling.

\subsection{Natural labeler}\label{sec:natural-labeler}

\begin{figure}
\centering
\includegraphics[width=0.8\textwidth]{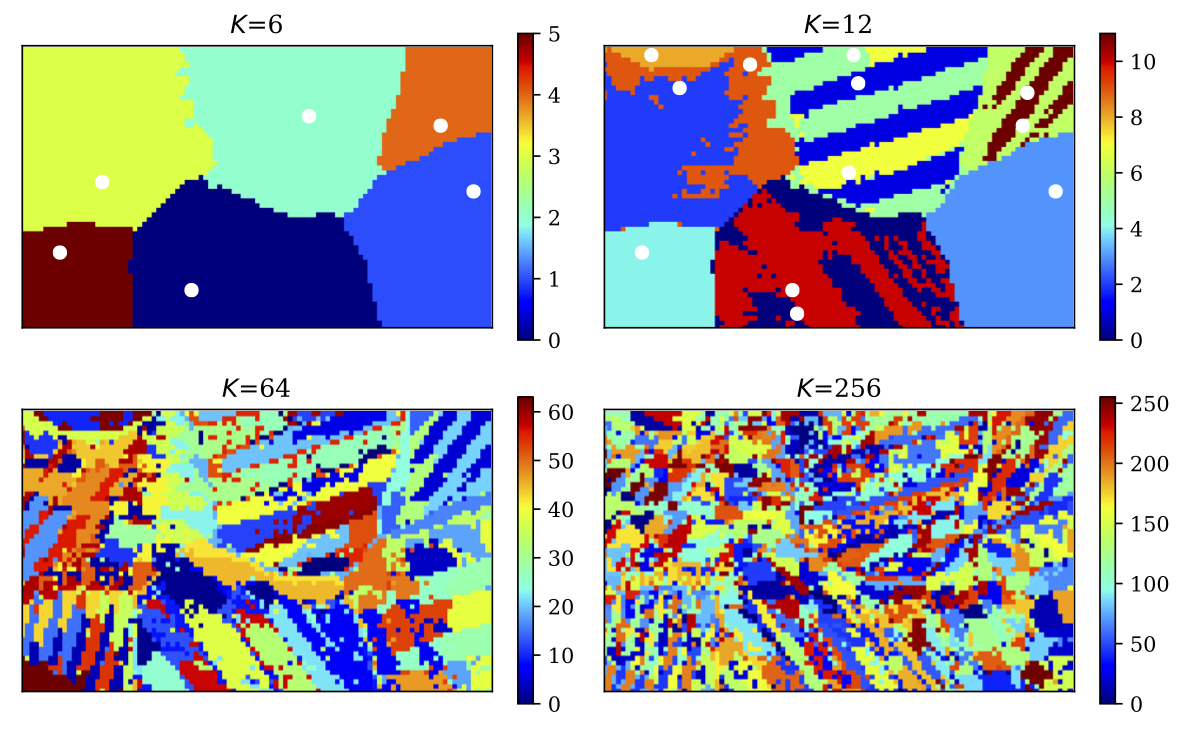}
\caption{Bayesian Gaussian Mixture of BaTiO$_3$ clustering colored by natural labeler. $M_{\rm PCA} = M_{\rm AE} = 256$. 
White dots are the delegate  points, which is omitted in the $K=64$ and $K=256$ cases.}
\label{fig:BGM_Natural}
\end{figure}

\begin{figure}
\centering
\includegraphics[width=0.8\textwidth]{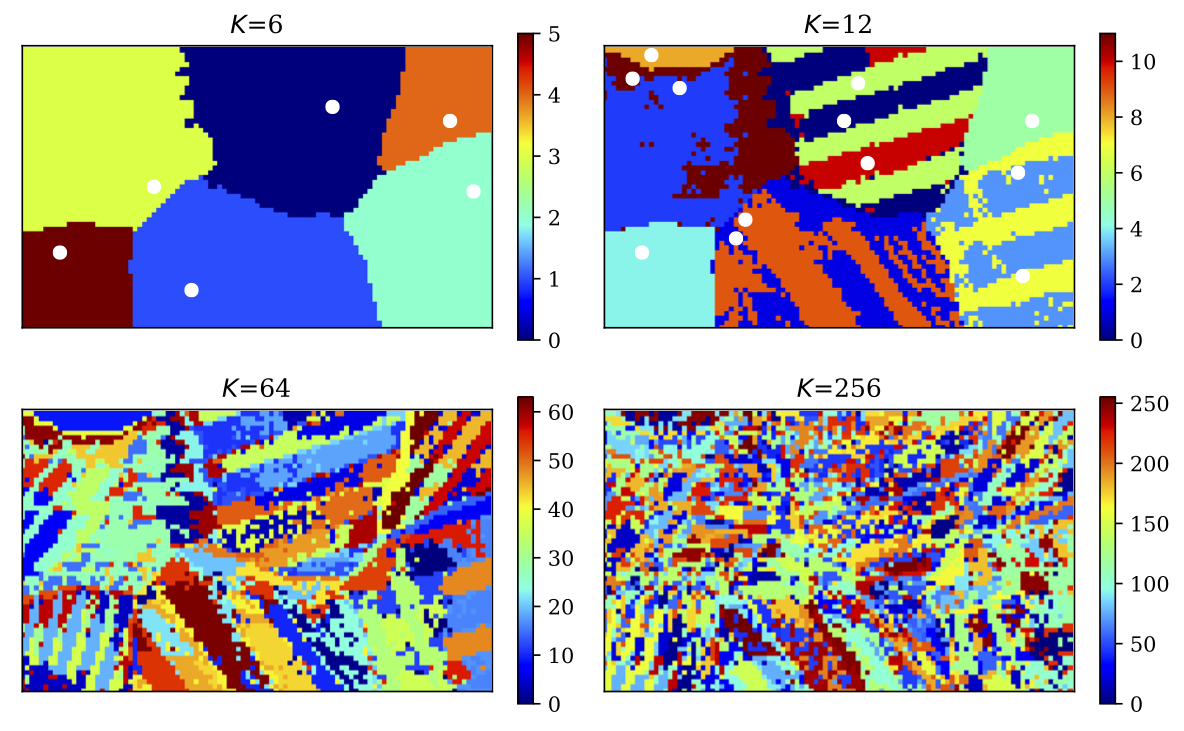}
\caption{K-Means clustering of BaTiO$_3$ with truncated principal components, colored by natural labeler. $M_{\rm PCA} = 32$ and $M_{\rm AE} = 256$. 
White dots are the delegate  points, which is omitted in the $K=64$ and $K=256$ cases.}
\label{fig:KM_Natural_32}
\end{figure}

\begin{figure}
\centering
\includegraphics[width=0.8\textwidth]{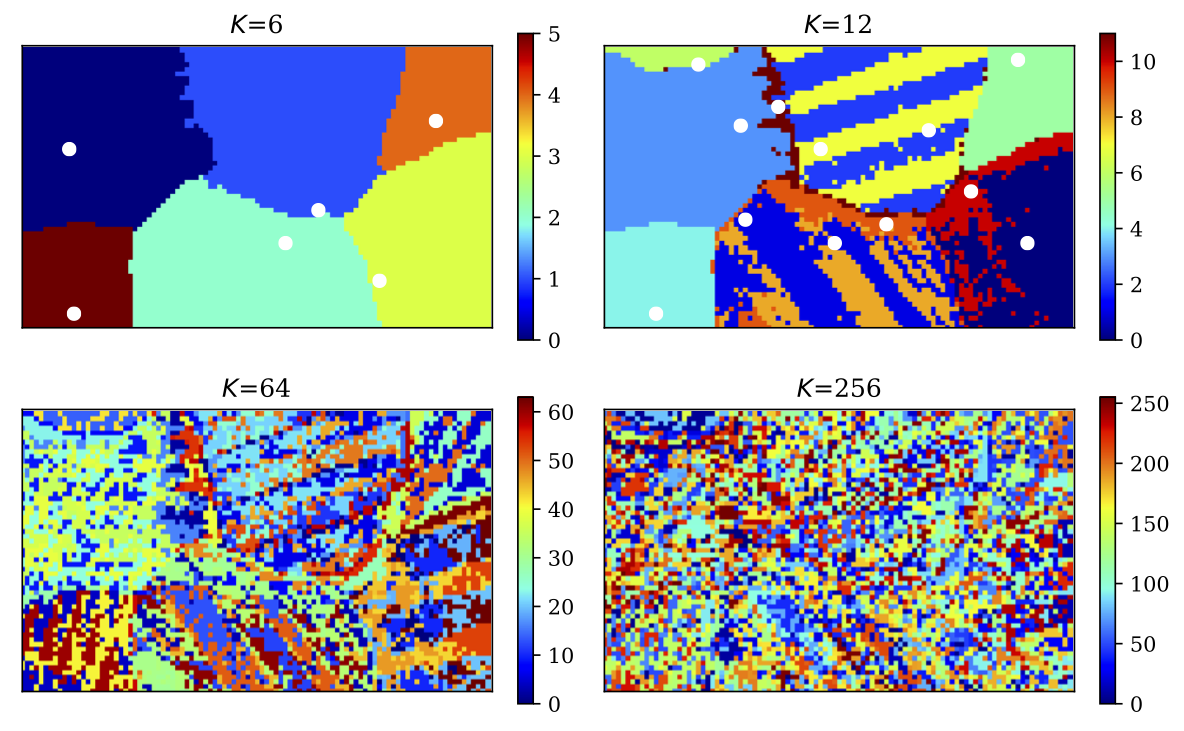}
\caption{K-Means clustering of BaTiO$_3$ with truncated principal components, colored by natural labeler. $M_{\rm PCA} = 8$ and $M_{\rm AE} = 256$. 
White dots are the delegate  points, which is omitted in the $K=64$ and $K=256$ cases.}
\label{fig:KM_Natural_8}
\end{figure}

Before going into metric analysis, we inspect the raw output of the segmentation.
Recall the natural labeler of $K$ clusters is the identity map $\ell_{\rm N}(k) = k$.
We plot the label map $\phi[v, g, \ell_{\rm N}]$ for various choices of parameters, in Figure \ref{fig:KM_Natural}, \ref{fig:BGM_Natural}, \ref{fig:KM_Natural_32}, \ref{fig:KM_Natural_8}. 
From these figures, we observe that:

\begin{figure}
\centering
\includegraphics[width=0.96\textwidth]{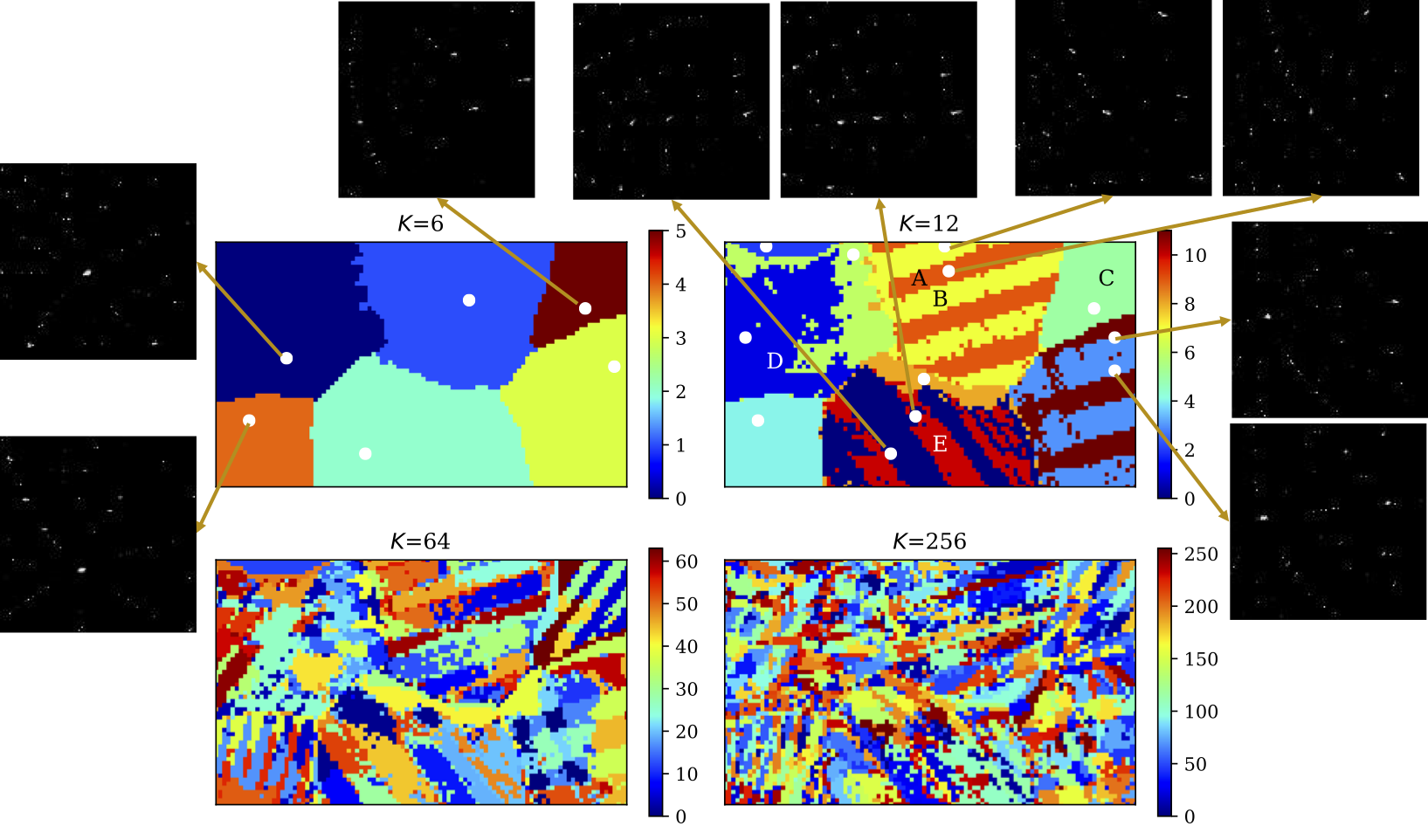}
\caption{K-Means clustering of BaTiO$_3$ colored by natural labeler. $M_{\rm PCA} = M_{\rm AE} = 256$. 
White dots are the delegate points, which is omitted in the $K=64$ and $K=256$ cases. 128$\times$128 pixel patterns of several delegate points are also given.}
\label{fig:KM_Natural}
\end{figure}

\begin{enumerate}
    \item At $K=6$, all segmentation models clearly partition the specimen domain into 6 grains. 
    \item At $K=12$, all segmentation models start to reveal the twinning microstructure in some grains.
    \item As $K$ increases, more and more fine features are extracted. However when $K$ is too large (e.g. $K = 64$ and $K = 256$), the segmentation colored by the natural labeler becomes mosaic. 
    \item No visible improvement or difference is observed in BGM segmentation (Figure \ref{fig:BGM_Natural}) compared to K-Means segmentation (Figure \ref{fig:KM_Natural}). \label{obs:BGM}
    \item No visible degradation or difference is observed in the K-Means segmentation using only the first 32 principal components (Figure \ref{fig:KM_Natural_32}) compared with K-Means using all features (Figure \ref{fig:KM_Natural}). However, K-Means corresponding to $M_{\rm PCA} = 8$ (Figure \ref{fig:KM_Natural_8}) is significantly noisier than that corresponding to $M_{\rm PCA}$. \label{obs:pca}
\end{enumerate}

By the observation \ref{obs:BGM} and \ref{obs:pca}, we will only consider K-Means with all 256 features and K-Means with the first 32 principal components in following discussions, unless otherwise mentioned.

\subsection{Unsupervised labeling}

\begin{figure}
\centering
\includegraphics[width=0.95\textwidth]{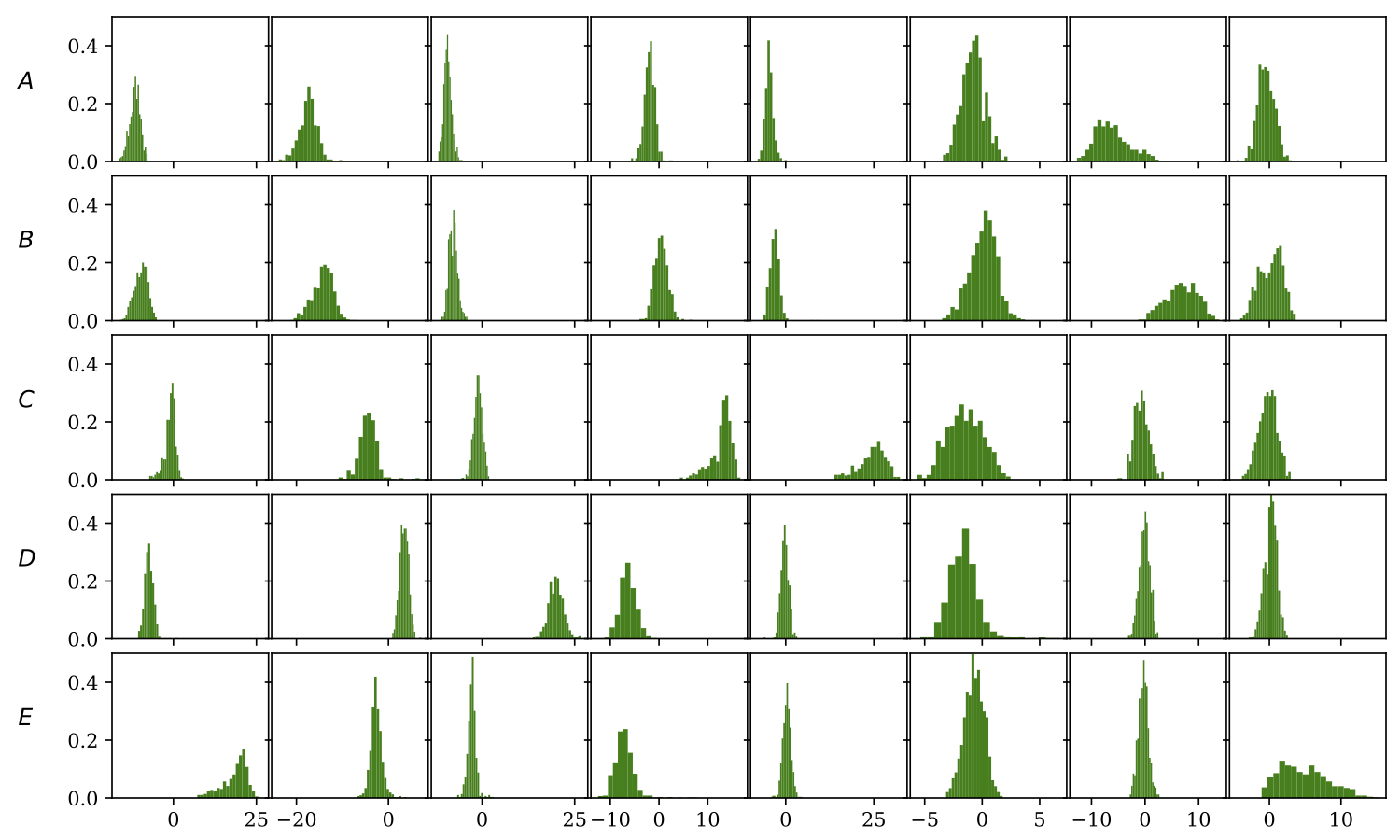}
\caption{Density distribution of the first 8 principal components for 5 clusters in the $K=12$ case in Figure \ref{fig:KM_Natural}.}
\label{fig:distribution}
\end{figure}

When the true orientation map is not available, $e.g.$ the X-ray crystallography software fails to
analyze portion or all of the Laue patterns from experiments, an unsupervised labeling becomes
necessary.
In order to plot the segmentation result, we need to define a labeler that is independent of the true map. A careful observation of the density distribution of clusters in the feature space (Figure \ref{fig:distribution}), guides us intuitively to conclude the following from the characteristics of clusters $A$, $B$, $C$, $D$ and $E$ in the $K=12$ case in Figure \ref{fig:KM_Natural}:
1) The shape of the clusters in the feature space are close to ellipsoids, that is the density distribution of each feature is close to a Gaussian.
2) Clusters $A$ and $B$ as a twinning pair in the same grain, has almost the same distribution, except for the $7$-th principal component, while the difference from either of them to $C$, $D$ or $E$ is significant.
The second observation demonstrates that the distance between patterns in the features space has a quantitative and sensitive correlation to the physical difference between their underlying crystal structures and crystallographic orientations.

\begin{figure}
\centering
\includegraphics[width=0.7\textwidth]{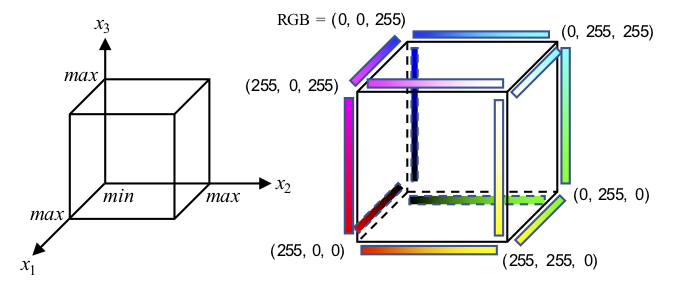}
\caption{Color scheme of the PCA labeler. Scale the range of the first 3 principal components, denoted $x_1$, $x_2$ and $x_3$, to the range of the red, green and blue channel of a RGB color code respectively.}
\label{fig:pca_cmap}
\end{figure}

\begin{figure}
\centering
\includegraphics[width=0.7\textwidth]{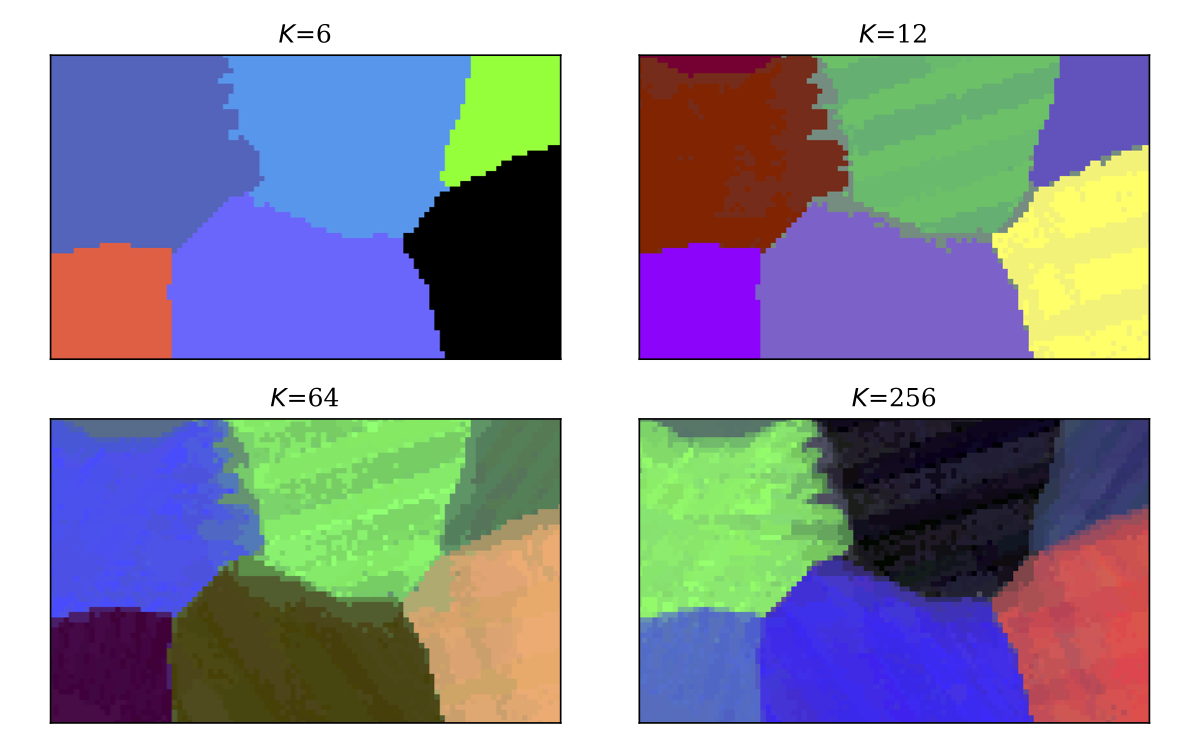}
\caption{K-Means clustering of BaTiO$_3$ colored by PCA labeler. $M_{\rm PCA} = M_{\rm AE} = 256$.
The coloring scheme is explained in the text and Figure \ref{fig:pca_cmap}.}
\label{fig:KM_PCA}
\end{figure}

\begin{figure}
\centering
\includegraphics[width=0.7\textwidth]{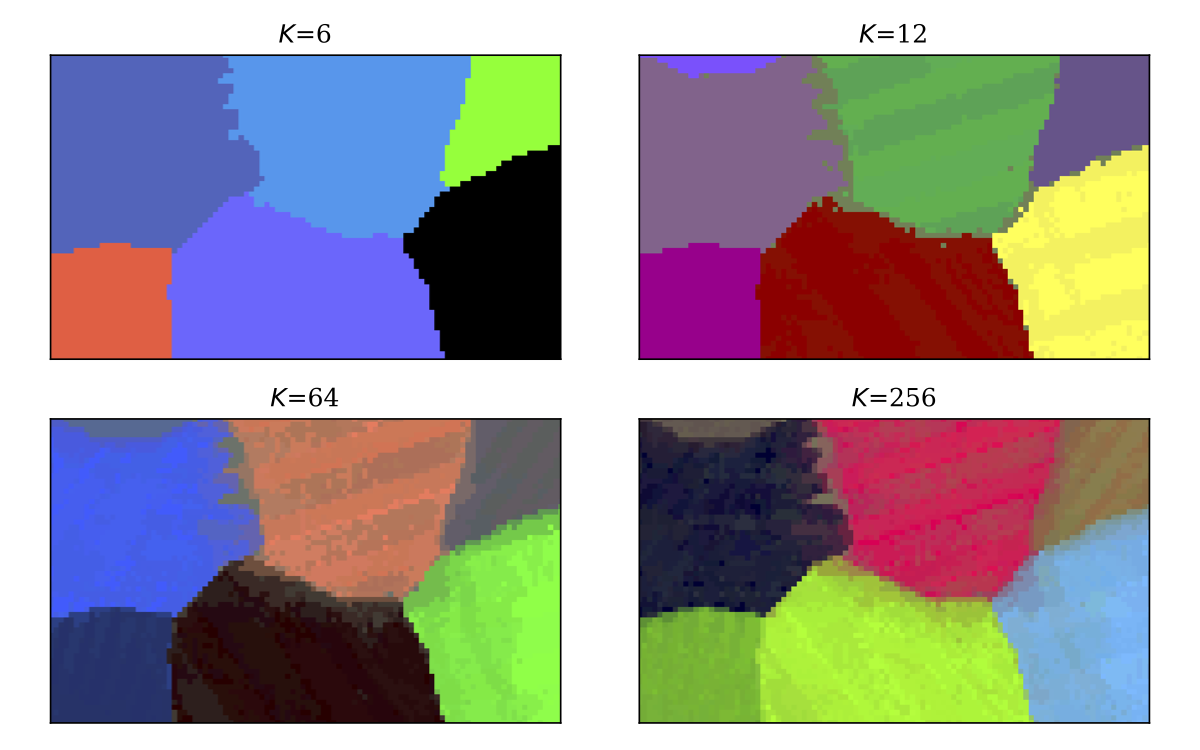}
\caption{K-Means clustering of BaTiO$_3$ with truncated principal components, colored by PCA labeler. $M_{\rm PCA} = 32$ and $M_{\rm AE} = 256$. 
The coloring scheme is explained in the text and Figure \ref{fig:pca_cmap}.}
\label{fig:KM_PCA_32}
\end{figure}

First, we use a PCA transformation from ${\mathbb R}^{M_{\rm PCA}}$ to $\mathbb R^3$ to transform the centroids of all the clusters $\{\mathbf u_k\}$ and truncate them to the first 3 principal components.
Subsequently, we scale the range of each projected components to $[0, 255]$. 
The color scheme is illustrated in Figure \ref{fig:pca_cmap}.
The composition of such a PCA transformation and the linear scaling is defined as the \emph{PCA labeler}, $\ell_{\rm PCA}$. 
$\ell_{\rm PCA}$ has 3 channels.
Thus, we define a coloring scheme that uses the 3 channels as the intensity of red, blue, and green channels respectively.
The results corresponding to the segmenations in Figure \ref{fig:KM_Natural} and \ref{fig:KM_Natural_32} are shown in Figure \ref{fig:KM_PCA} and \ref{fig:KM_PCA_32} respectively.

The maps for large $K$ ({\it e.g.} $K=64$ and $K=256$) colored by the PCA labeler are much more comprehensible than their counterpart colored by the natural labeler.
More importantly, the fine features in the map are stable as $K$ increases.
It suggests that without concerning the subsequent indexing effort, the more clusters used in the segmentation, the better the pipeline extracts the spatial features in the specimen domain.
Pushing this thought to an extreme, we set $K = N$. 
That is each sample is a cluster by itself.
In other words, we directly use the first 3 principal components to color all feature samples, without any clusters. 
This leads to the following definition.
\begin{definition}
The pipeline $(v_{\rm PCA}, g_I, \ell_{\rm PCA})$ is called the \emph{direct coloring} and the associated segmentation $(v_{\rm PCA}, g_I)$ is the \emph{direct segmentation},
where the dummy estimator $g_I: v_{\rm PCA}({\cal S}) \to \{1, ..., N\}$ assigns each of the feature samples to an individual cluster.
\end{definition}

\begin{figure}
\centering
\includegraphics[width=0.35\textwidth]{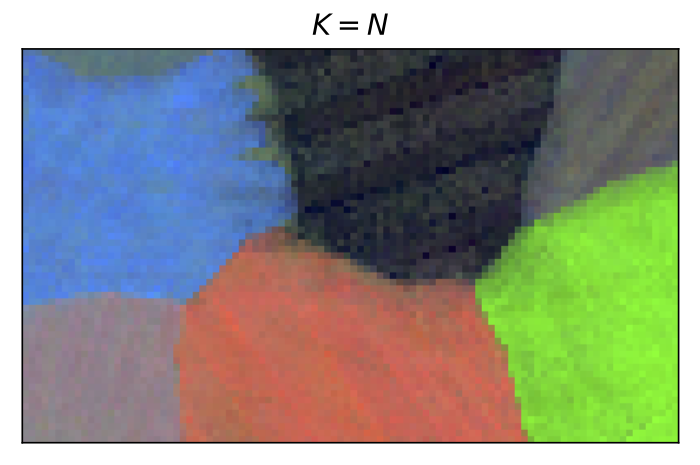}
\caption{
Direct coloring of BaTiO$_3$ ($M_{\rm AE} = 256$).
The coloring scheme is explained in the text and Figure \ref{fig:pca_cmap}.
\label{fig:KM_PCA_N}
}
\end{figure}

The result of applying the direct coloring to BaTiO$_3$ is shown in Figure \ref{fig:KM_PCA_N}.
The direct segmentation $(v_{\rm PCA}, g_I)$ retains the most complete information resulting from the feature map $v_{\rm PCA}$.
Any non-trivial clustering $(v_{\rm PCA}, g)$ causes certain information loss.
When the conventional indexing procedure is either not available or not needed for the experiment, the direct segmentation $(v_{\rm PCA}, g_I)$, {\it i.e.} pre-clustering, with an appropriate labeler can be used to analyze the X-ray microdiffraction scans independent of any crystallographic information of the material.
The PCA labeler as shown in Figure \ref{fig:KM_PCA_N} is a good clustering tool in general.
Even in the scenarios where we have to group feature samples into a small number of clusters,
we can still use the direct segmentation as a reference to help choosing $K$.

\begin{figure}
\centering
\includegraphics[width=0.8\textwidth]{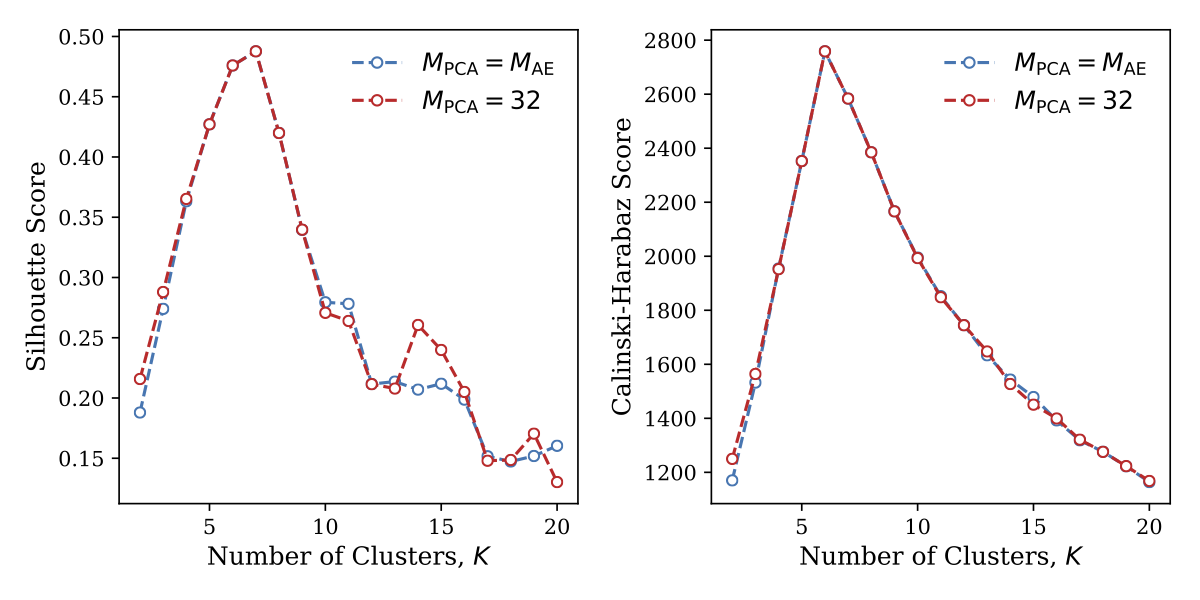}
\caption{Silhouette score and Calinski-Harabaz score for K-Means segmentation of BaTiO$_3$ with different $K$.}
\label{fig:SCH}
\end{figure}

In all above segmentation models, $K=6$ consistently gives the clearest segmentation of grains. 
This is because the difference in the diffraction patterns across grains is much larger than that within a grain. 
We say this specimen has 6 \emph{major subdomains}.
In most of the materials, the major subdomains are separated by grain boundaries or phase boundaries.
Figure \ref{fig:SCH} plots the Silhouette score \cite{rousseeuw1987silhouettes} and the Calinski-Harabaz score \cite{calinski1974dendrite} for $K$ between 2 and 20.
Peaks in both scores occur near $K=6$, which explains the aforementioned observation about major subdomains.

\subsection{Supervised labeling}

\begin{figure}
\centering
\includegraphics[width=0.7\textwidth]{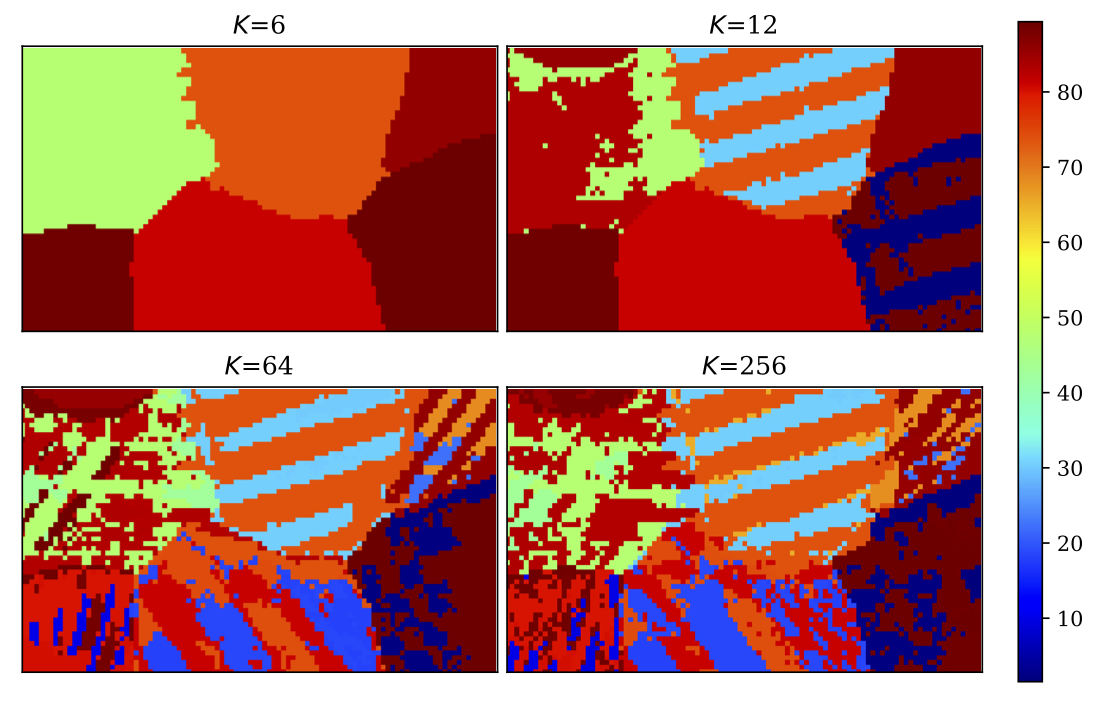}
\caption{K-Means clustering of BaTiO$_3$ colored by indexing labeler. $M_{\rm PCA} = M_{\rm AE} = 256$.}
\label{fig:KM_Indexing}
\end{figure}

\begin{figure}
\centering
\includegraphics[width=0.7\textwidth]{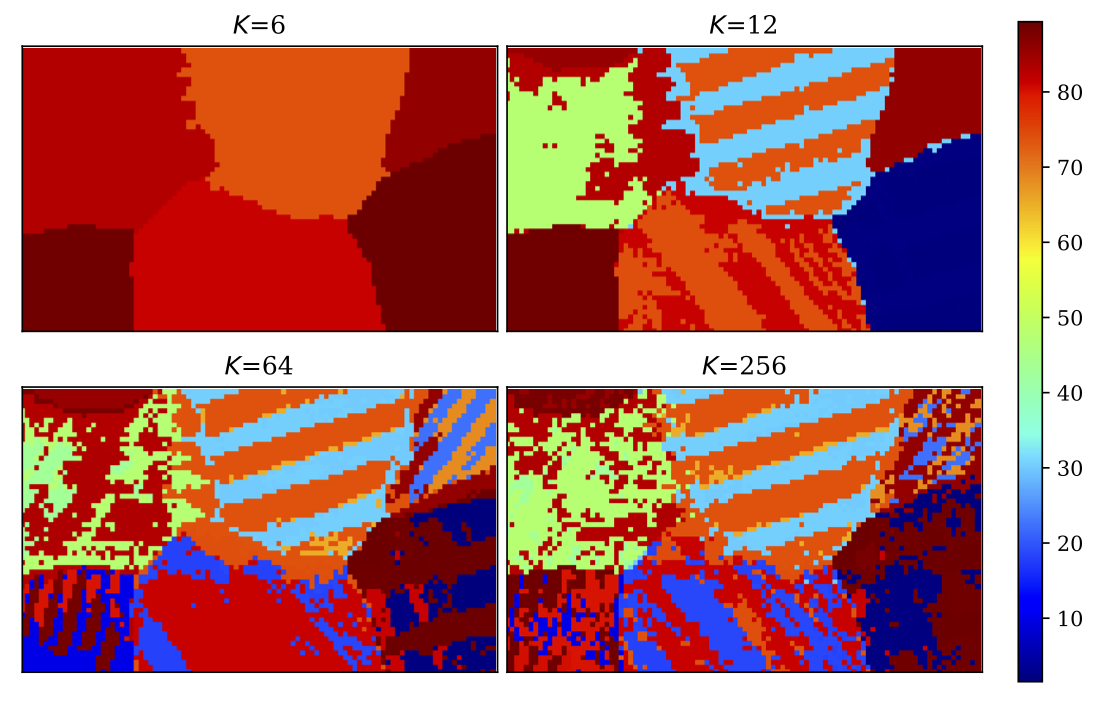}
\caption{K-Means clustering of BaTiO$_3$ with truncated principal components, colored by indexing labeler. $M_{\rm PCA} = 32$ and $M_{\rm AE} = 256$.} 
\label{fig:KM_Indexing_32}
\end{figure}

In a supervised labeling, we utilize the true orientation map as a result of complete indexing to assess the quality of a segmentation pipeline $\{v, g\}$.
Denote the true orientation map, as shown in Figure \ref{fig:BTO}, as $o(x,y)$.
\begin{definition}
The \emph{indexing labler} of a property map $f(x, y)$ gives the true value of $f$ at the delegate  point $w(k)$ for the cluster ${\cal C}_k$:
\begin{equation}
    \ell_{\rm I}[f](k) = f(w(k)).
\end{equation}
\end{definition}
When assessing a segmentation model via supervised method, we can directly use $o(x, y)$ to evaluate $\ell_{\rm I}[o]$.
In practice, to get $\ell_{\rm I}[o]$ one needs to get the true value of orientation at the delegate  points $\{(x_k, y_k)\}$ by indexing only the patterns at those points.
Appending the indexing labler to a segmentation $(v, g)$ completes a pipeline $(v, g, \ell_{\rm I}[o])$, and therefore generates a label map $\phi[v, g, \ell_{\rm I}[o]]$.
We plot in Figure \ref{fig:KM_Indexing}, \ref{fig:KM_Indexing_32} the same segmentations as in Figure \ref{fig:KM_Natural}, \ref{fig:KM_Natural_32}, and colored by the indexing labler.

\begin{figure}
\centering
\includegraphics[width=0.8\textwidth]{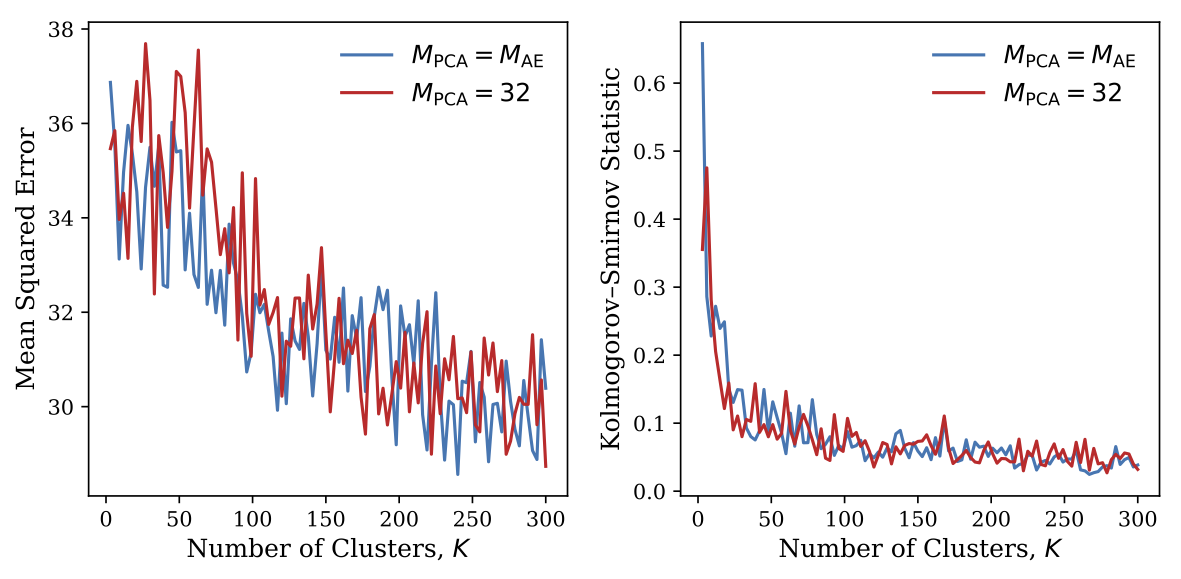}
\caption{Mean squared error and Kolmogorov-Smirnov distance between the clustered approximation by K-Means and the true orientation map.}
\label{fig:se}
\end{figure}

To assess the quality of the segmentation $(v, g)$, we can check how well $\phi[v, g, \ell_{\rm I}[o]]$ approximate the true map $o(x, y)$. 
A natural metric for the latter check is the mean squared error between the two property maps .
A less accurate but computationally cheaper test is the Kolmogorov-Smirnov distance, computed by the two-sample Kolmogorov-Smirnov test, between the two data sets $o({\cal S})$ and $\phi[v, g, \ell_{\rm I}[o]]({\cal S})$.
Figure \ref{fig:se} shows the mean squared error and the Kolmogorov-Smirnov distance for a wide range of $K$.
While the mean squared error in general decreases as $K$ increases, the reduction is very slow when $K$ is large, {\it e.g.} when $K > 100$.
This behavior is more prominent when it is studied by the Kolmogorov-Smirnov distance.
This suggests that one can get an approximation with almost the same quality even using a much less number of clusters, {\it i.e.} much less number of patterns to index.
Indeed, using just 64 or 256 delegate  patterns, we get an orientation map that is fairly close to the true map resulting from indexing 6000 patterns.
Nevertheless, the exact tolerant of approximation error depends on the specific physical problem of study. 
If the tolerance turns out to be small, we need to use a large $K$ to achieve a higher accuracy.

\subsection{Pre-Index Segmentation Procedure}

Before heading to more examples, we summarize the common procedure of pre-index segmentation
\begin{enumerate}
    \item Pick a feature extractor. Normally a CNN autoencoder with $M_{\rm AE}$ latent features. As we are going to see in the examples, in most cases, $M_{\rm AE} = 256$ is sufficient. In certain experiments, we might need a more complex extractor.
    \item Pick a PCA transformation with $M_{\rm PCA} = M_{\rm AE}$ to further reduce the dimension of the feature space for the particular experiment under study. This completes the feature map $v_{\rm PCA}$. 
    \item Plot direct coloring for $v_{\rm PCA}$. The direct coloring gives us the first overview of the spatial features in the specimen. In some cases, the direct coloring itself can already help us advance in the scientific study. The first 3 steps are quite standard for all Laue scan experiments. 
    In general, we should decide the next steps based on the direct coloring and the particular physical problem to be solved. The following are the common options.
    \item Plot Silhouette score and Calinski-Harabaz score for several $K$ values. Try to identify the major subdomains with the scores. Recall that major subdomains are usually separated by the grain boundaries and phase boundaries, which are the first-order microstructural features that are crucial in most experiments.
    \item Visually identify some potentially important microstructural features from the direct labeling. Manipulating $K$ to retrieve such features in label map using a reasonably small $K$ value.
    \item If the required $K$ turns out to be big, and therefore the clustering algorithm is too slow with all features, revisit Step 2 above, further truncate the principal components under the guidance of the distribution of explained variance of principal components.
    \item Finally if needed and feasible, index the $K$ delegate  patterns. Then use the indexing labeler to get the approximated property map.
    \item There is a useful strategy, which is not illustrated in the examples in later section:
    one can perform a coarse scan on a large area first, and get the direct coloring.
    By looking at the direct labeling, one may be able to locate the area of interest where a dense scan in a small area can be performed.
\end{enumerate}

\section{Examples}

In this section we investigate 3 additional examples using the pre-index segmentation algorithm to
analyze the Laue microdiffraction data collected at the Advanced Light Source Beamline 12.3.2,
Lawrence Berkeley National Lab. Experimental details on the data collection procedure and description
of the beamline apparatus have been published elsewhere \cite{tamura2003scanning, tamura2014xmas}. Unless mentioned otherwise, we will use the feature map of $M_{\rm PCA} = M_{\rm AE} = 256$.

\subsection{CuAlMn alloy}

\begin{figure}
\centering
\includegraphics[width=0.9\textwidth]{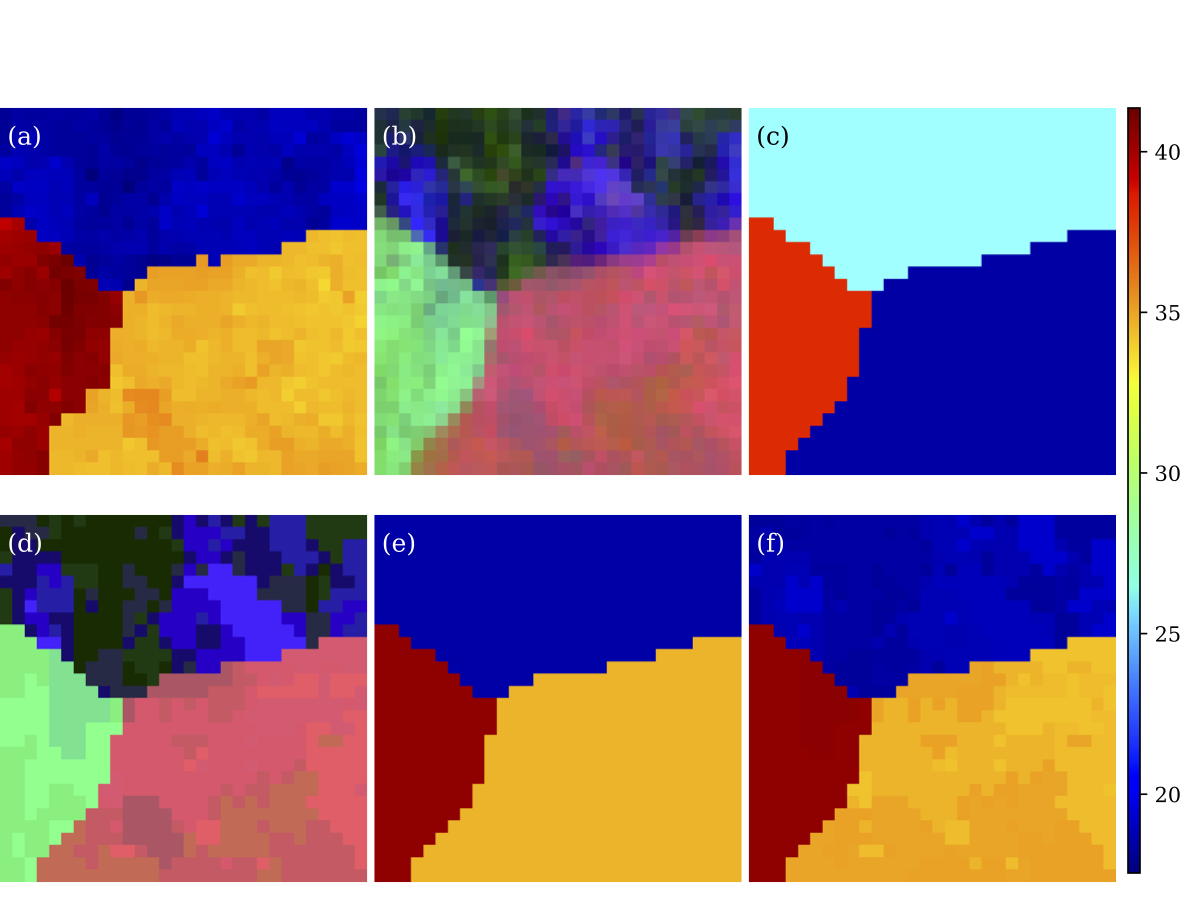}
\caption{In CuAlMn specimen:
(a) True map. 
(b) Direct coloring. 
(c) PCA labeler with $K=3$.
(d) PCA labeler with $K=16$.
(e) Indexing labeler with $K=3$.
(f) Indexing labeler with $K=16$.
The color bar on the right is for (a)(e)(f).
The coloring scheme for (b)(c)(d) is explained in the text and Figure \ref{fig:pca_cmap}.}
\label{fig:C5}
\end{figure}

\begin{figure}
\centering
\includegraphics[width=0.8\textwidth]{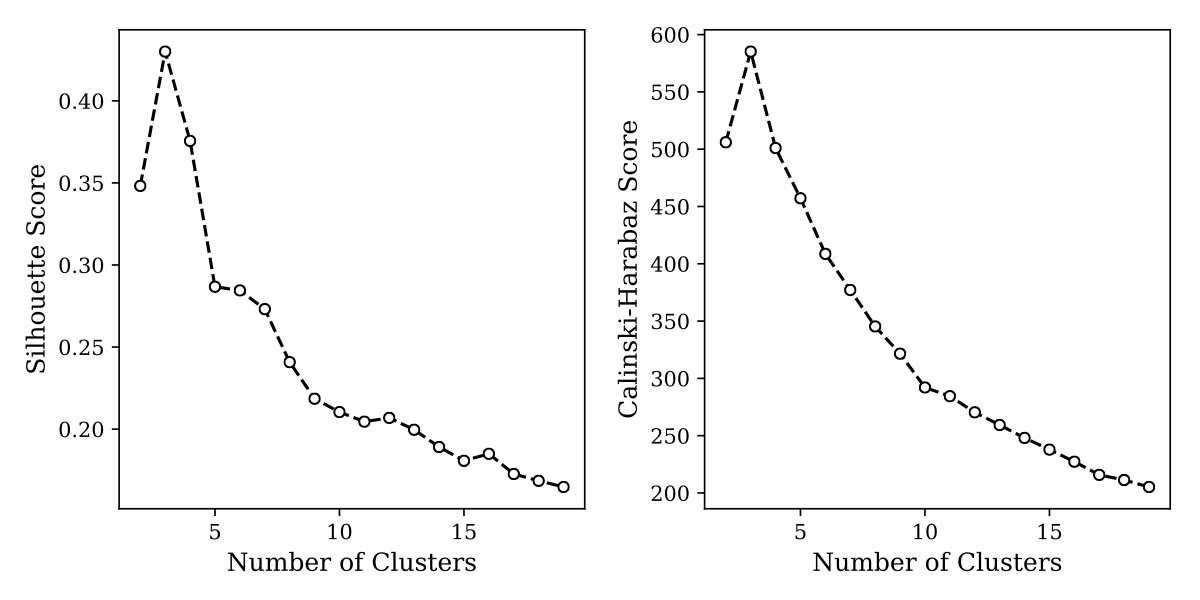}
\caption{Silhouette score and Calinski-Harabaz score for CuAlMn specimen with different $K$.}
\label{fig:C5_SCH}
\end{figure}

This alloy with atomic composition 70.8\% Cu -- 21.4\% Al -- 7.8\% Mn undergoes a martensitic phase
transformation at -7$^\circ$C. We conducted the synchrotron microdiffraction scan at room temperature
on an area consisting of several austenite grains. The austenite crystal structure can be well
indexed by symmetry Fm$\bar3$m, and we use the XMAS parallel analysis algorithm to sequentially
index the whole data set and generate the true orientation map in Figure \ref{fig:C5}a. This example
shows that the pre-index segmentation algorithm correctly captures the microstructure of a single
phase, multi-grains material regardless of the types of labeler and number of clusters. In the true
orientation map, there exist some points, especially those near the grain boundaries where the Laue
patterns failed to be indexed. By pattern segmentation, these points can be identified, and the grain
boundaries are better resolved. In addition, the scanned area consists of three main clusters based
on the Silhouette and Calinski-Harabaz score (Figure \ref{fig:C5_SCH}). By increasing the number of clusters,
the detailed microstructures in each of the main grains can be resolved better, e.g. Figure \ref{fig:C5}b,
d and f.

\subsection{AuCuZn alloy}

\begin{figure}
\centering
\includegraphics[width=0.9\textwidth]{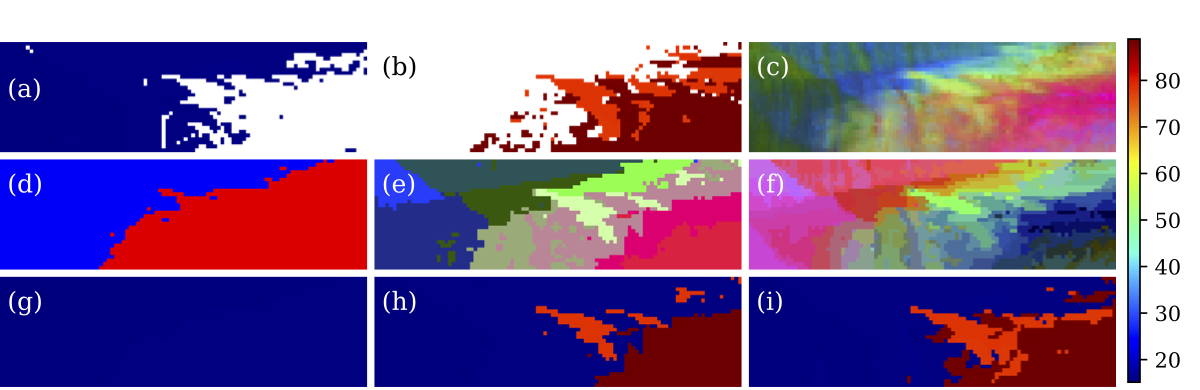}
\caption{In Au$_{29}$Cu$_{26}$Zn$_{45}$ alloy:
(a) True map for austenite. 
(b) True map for martensite. 
(c) Direct coloring.
(d)(e)(f) PCA labeler with $K=2$, 10 and 64.
(g)(h)(i) Indexing labeler with $K=2$, 10 and 64.
The color bar on the right is for (a)(b)(g)(h)(i).
The coloring scheme for (c)(d)(e)(f) is explained in the text and Figure \ref{fig:pca_cmap}.
}
\label{fig:Au29}
\end{figure}

\begin{figure}
\centering
\includegraphics[width=0.8\textwidth]{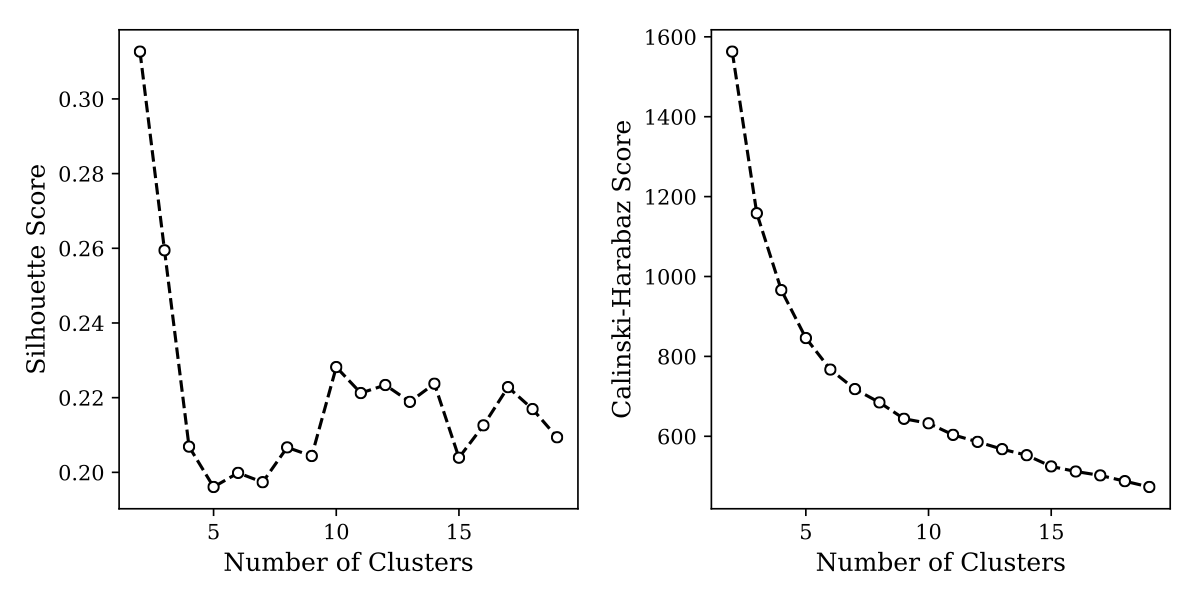}
\caption{Silhouette score and Calinski-Harabaz score for Au$_{29}$Cu$_{26}$Zn$_{45}$ alloy with different $K$.}
\label{fig:Au29_SCH}
\end{figure}

The phase-transforming alloy Au$_{29}$Cu$_{26}$Zn$_{45}$ is an example having the complex interface morphology between austenite and martensite phases. At around -40$^\circ$C, the cubic austenite transforms to the monoclinic martensite.
To generate the true orientation map, we have to run the indexing algorithm twice for the same scanned area: one for cubic symmetry and the other for monoclinic symmetry corresponding to the Figure \ref{fig:Au29}a and b respectively.
By checking Silhouette score and Calinski-Harabaz score (Fig.\ref{fig:Au29_SCH}), there are only 2 major subdomains. 
A PCA labeling of $K=2$ clustering clearly reveals the phase boundary (Figure \ref{fig:Au29}d)
But the corresponding indexing labeler fails to resolve the boundary clearly (Figure \ref{fig:Au29}g) because the delegate selected for the whole martenstite cluster is failed to be indexed and can not represent the martensite region consisting of the martensite variants with different orientations. 
When we increase $K$ to 10 and 64,
both the PCA labeler (Figure \ref{fig:Au29}e, f) and the indexing labeler (Figure \ref{fig:Au29}h, i) reveals finer and richer microstructures of martensite.
The indexing labeler at large number of clusters starts converging to the true map.

\subsection{CuAlNi alloy}

\begin{figure}
\centering
\includegraphics[width=0.9\textwidth]{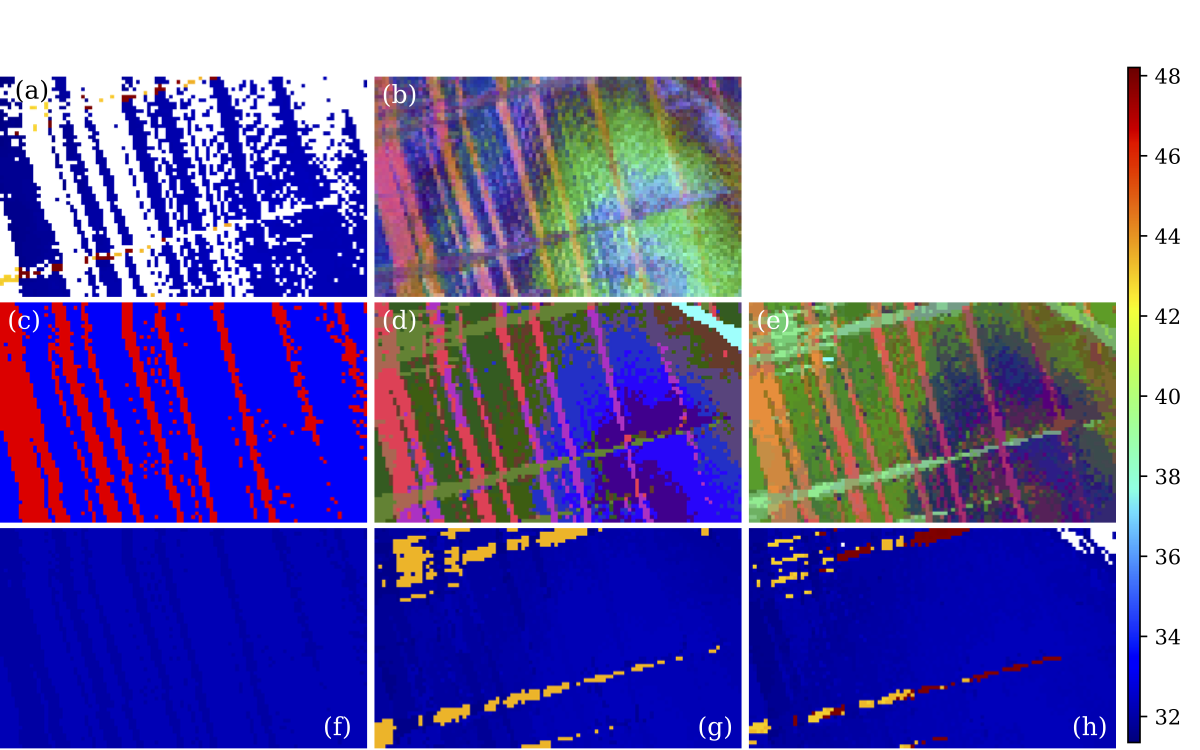}
\caption{In CuAlNi alloy:
(a) True map. 
(b) Direct coloring.
(c)(d)(e) PCA labeler with $K=2$, 12 and 64.
(f)(g)(h) Indexing labeler with $K=2$, 12 and 64.
The color bar on the right is for (a)(f)(g)(h).
The coloring scheme for (b)(c)(d)(e) is explained in the text and Figure \ref{fig:pca_cmap}.
}
\label{fig:CAN}
\end{figure}

\begin{figure}
\centering
\includegraphics[width=0.8\textwidth]{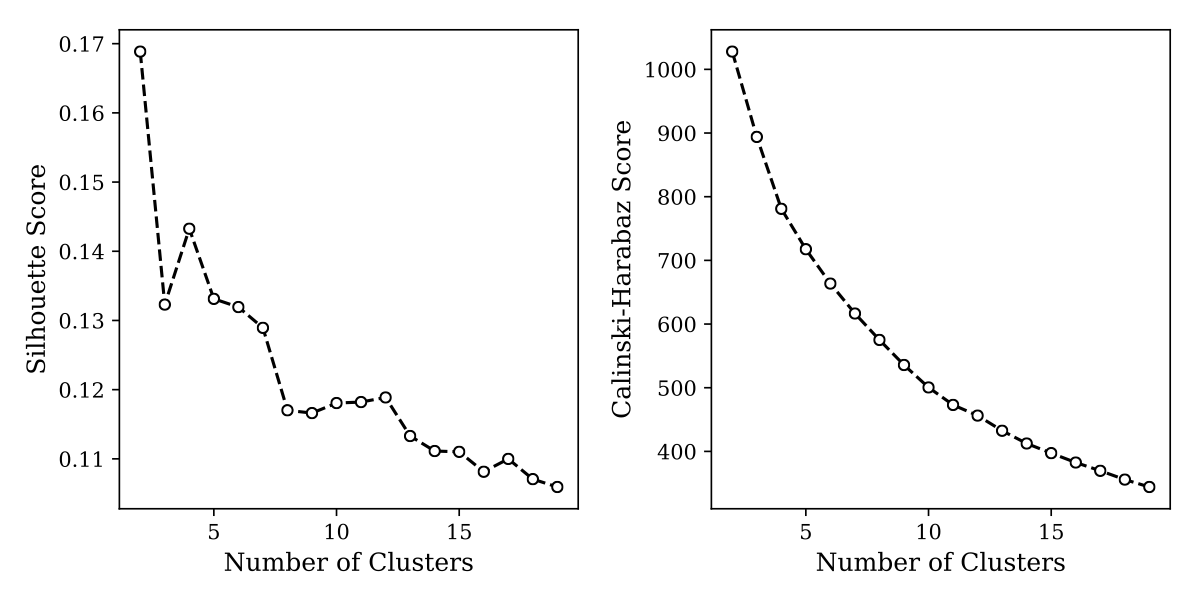}
\caption{Silhouette score and Calinski-Harabaz scores for CuAlNi with different $K$.}
\label{fig:CAN_SCH}
\end{figure}

CuAlNi alloy is in martensite phase made up of fine twins.
Since the fineness of some of the twinning structure is beyond the resolution of the X-ray microdiffraction, the indexing was not very successful. As shown in Figure \ref{fig:CAN}a, There are a large area failed to be indexed. 
One can only roughly see certain vertical features that might correspond to the twins. In this case, the direct coloring is strongly preferred because of the lack of true map. 
The segmentation result by the direct coloring is shown in Figure \ref{fig:CAN}b, which clearly reveal the twin feature of the specimen.
The Silhouette score and Calinski-Harabaz score (Figure \ref{fig:CAN_SCH}) suggests 2 major subdomains. 
Figure \ref{fig:CAN}c is $K=2$ clustering colored by the PCA labeler, which shows clear twin boundaries.
In contrast, because of poor indexing, the indexing labeling (Figure \ref{fig:CAN}f) is not applicable in this case.
As we increase $K$ to 12 and 64,
the PCA labeler (Figure \ref{fig:CAN}d, e) starts to show finer and finer features of twinning microstructures.

\section{Real-time Performance}

As stated earlier, one of the motivations of the index-free segmentation is to enable real-time
analysis of Laue XRD data. To justify the feasibility, we list performance profile in Table \ref{tbl:benchmark}. 
In a real application, feature extraction can be done on a streaming fashion: extract features from a
pattern right after it is taken by the detector. Then direct coloring and some other label maps can
be computed immediately. In contrast, the average time accounted for the indexing process with known structural parameters varies from 10 seconds to several minutes. For a $30 \times 30$ domain, the total analysis time by the traditional method would be several hours to days.
If further tuning is needed, interactive re-clustering is possible, given
the speed of computation.

\begin{table}
\label{tbl:benchmark}
\begin{threeparttable}
\caption{Performance of each phase in a pre-index segmentation pipeline.\tnote{1}}
\begin{tabular}{cccc}
\hline
Study Case & Data Size & Time for Feature Extraction (sec)\tnote{2} & Time for Clustering (sec)\tnote{3} \\
\hline
CuAlMn & $30\times30$ & 45 & 2.3 ($K=3$); 5.6 ($K=16$) \\
AuCuAn & $30\times100$ & 63 &  2.7 ($K=2$); 6.7 ($K=10$); 14.3 ($K=64$) \\
CuAlNi & $60\times100$ & 104 &  3.5 ($K=2$); 12.7 ($K=12$); 23.4 ($K=64$) \\
\hline
\end{tabular}
{\small
\begin{tablenotes}
    \item[1] Measured in a desktop environment with Intel i7-8700K (6 physical cores at 3.7GHz) CPU and 16GB RAM.
    \item[2] Using 12 parallel jobs.
    \item[3] The numbers in the parenthesis are $K$ (number of clusters) values.
  \end{tablenotes}
}
\end{threeparttable}
\end{table}

\section{Conclusion}

In this paper, we proposed a machine learning based data processing pipeline for synchrotron Laue
X-ray microdiffraction experiments. We formalized the pipeline to consist of 3 phases: feature extraction,
clustering, and labeling. We then demonstrated the procedure of getting an approximated
property map from Laue patterns in 4 different examples with distinct types of materials. The results
are promising and the performances are impressive. A real-time data processing platform
could be built on top of this pipeline. This approach can be easily extended to electron diffraction based characterization such as EBSD, where the Kikurchi pattern plays the same role as the Laue pattern. Some deep learning models \cite{jha2018extracting} have already established for the feature extraction of the EBSD pattern, which can be adopted to our clustering and labelling pipeline.

In fact, each of the phases in the pipeline can be further studied separately. For feature extraction, other CNN architectures
can be explored. One could also try other feature extraction methods with more direct
physical meaning, such as using the complete list of peak positions, intensities and shapes. For clustering,
one could study other clustering algorithms, such as Gaussian Mixtures, DBSCAN, Mean
Shift, etc. Also, advanced image segmentation techniques can be utilized. For example, Markov
Random Field is the state of the art statistical model for image segmentation. For labeling, one
of the lesson learned from this paper is that a good labeler mapping the feature space into a low
dimensional space with dimension less than or equal to 3 greatly helps us to visualize the property
map, even without indexing. Finding better labelers, which includes finding the relationship between
hidden feature space and the true physical properties, should be a persistent goal for future
research.

\section*{Acknowledgement}
M. K. and X. C. thank the financial support of the HK Research Grant Council under GRF Grant 16207017 and 26200316. XC would like to thank the Isaac Newton Institute for Mathematical Sciences, Cambridge, for support and hospitality during the programme The mathematical design of new materials where work on this paper was undertaken. This work was supported by EPSRC grant no EP/K032208/1. Advaned Light Source was supported by the Office of Science, Office of Basic Energy Sciences, of the U.S. Department of Energy under Contract no. DE-AC02-05CH11231.

\bibliographystyle{naturemag}
\bibliography{segmentation}

\end{document}